\documentclass[a4paper,11pt]{article}
\usepackage{jheppub} 

\usepackage{amsfonts}
\usepackage{amsmath}
\allowdisplaybreaks[4]        
\usepackage{amssymb}
\usepackage{euscript}     
\usepackage{color}         
\usepackage{tensor}        
\usepackage{amsthm}
\usepackage{graphicx}
\usepackage{subfigure}
\usepackage{bbm}
\usepackage[header,title,page,titletoc]{appendix}  

\usepackage[numbers]{natbib}  
\usepackage{float}

\newcommand{\be}{\begin{equation}}
\newcommand{\bea}{\begin{eqnarray}}
\newcommand{\eea}{\end{eqnarray}}
\newcommand{\ba}{\begin{array}}
\newcommand{\ea}{\end{array}}
\newcommand{\ee}{\end{equation}}
\newcommand{\bes}{\begin{equation*}}
\newcommand{\beas}{\begin{eqnarray*}}
\newcommand{\eeas}{\end{eqnarray*}}
\newcommand{\bas}{\begin{array*}}
\newcommand{\eas}{\end{array*}}
\newcommand{\ees}{\end{equation*}}

\title{\boldmath On the role of counterterms in holographic complexity}

\author[a]{Amin Akhavan}
\author[b,1]{Farzad Omidi ,\note{Corresponding author.}}

\affiliation[a]{
Young Researchers Club, Centarl Tehran Branch, Islamic Azad University,
Tehran, Iran}
\affiliation[b]{School of Physics, Institute for Research in Fundamental Sciences (IPM), \\
P.O. Box 19395-5531, Tehran, Iran}

\emailAdd{amin$_{-}$akhavan@ipm.ir}
\emailAdd{farzad@ipm.ir}

\abstract{
We consider the Complexity=Action (CA) proposal in Einstein gravity and investigate new counterterms which are able to remove all the UV divergences of holographic complexity. We first show that the two different methods for regularizing the gravitational on-shell action proposed in ref. \cite{Carmi:2016wjl} are completely equivalent, provided that one considers the Gibbons-Hawking-York term as well as new counterterms inspired from holographic renormalization on timelike boundaries of the WDW patch. Next, we introduce new counterterms on the null boundaries of the WDW patch for four and five dimensional asymptotically AdS spacetimes which are able to  remove all the UV divergences of the on-shell action. Moreover, they are covariant and do not change the equations of motion. At the end, by applying the null counterterms, we calculate the holographic complexity of an AdS-Schwarzschild black hole as well as the complexity of formation. We show that the null counterterms do not change the complexity of formation.}


\arxivnumber{1906.09561}

\begin{document} 
	\maketitle
	\flushbottom
\section{Introduction}
The discovery of Ryu-Takayanagi \cite{Ryu:2006bv,Ryu:2006ef,Lewkowycz:2013nqa} and HRT formulas  \cite{Hubeny:2007xt,Dong:2016hjy} marked the beginning of a new era during which we have been encountering increasing evidence \cite{Ryu:2006bv,Ryu:2006ef,Hubeny:2007xt,Lewkowycz:2013nqa,Dong:2016hjy,Casini:2011kv,Nishioka:2009un,Hayden:2011ag,Takayanagi:2017knl,Nguyen:2017yqw,Umemoto:2018jpc,Caputa:2018xuf,Susskind:2014rva,Stanford:2014jda,Susskind:2014moa,Brown:2015bva,Susskind:2018pmk,Brown:2015lvg} that quantum gravity and quantum information theory are two indispensable topics. The connection is strong enough that one might introduce a dictionary between them and apply the AdS/CFT correspondence \cite{Maldacena:1997re} to calculate diverse quantities such as entanglement entropy \cite{Ryu:2006bv,Ryu:2006ef,Hubeny:2007xt,Casini:2011kv,Nishioka:2009un},  mutual information \cite{Hayden:2011ag,Allais:2011ys,Fischler:2012uv} and entanglement of purification \cite{Takayanagi:2017knl,Nguyen:2017yqw,Umemoto:2018jpc,Caputa:2018xuf}. 
\\Another important concept in quantum information theory  is computational complexity which is believed to be useful in understanding the interior of black holes \cite{Susskind:2014rva,Stanford:2014jda,Susskind:2014moa,Brown:2015bva,Susskind:2018pmk}.
In quantum field theory, computational complexity is defined as the minimum number of gates, i.e. simple unitary operations, needed to make a specific state from a reference state. For quantum states which are dual to black holes in AdS, computational complexity has interesting properties \cite{Susskind:2014rva,Susskind:2018pmk}: it is an extensive quantity and after the thermal equilibrium it increases linearly with time until it reaches its maximum value $e^S$, where $S$ is the thermal entropy of the black hole. Next,  it fluctuates around this value for a long time which is of order the quantum recurrence time $ e^{e^S}$, then it reduces to its minimal value. Moreover, generally the growth rate of complexity approaches the conjectured Lloyd bound \cite{Brown:2015bva,Lloyd:2000,Brown:2015lvg,Susskind:2018pmk,Cai:2016xho} from above at late times, and hence violates the bound \cite{Carmi:2017jqz,Kim:2017qrq,Couch:2017yil,Moosa:2017yiz,Alishahiha:2018tep}. However, for a one-sided black hole which is holographic dual to a CFT state after a global quench, the Lloyd bound is respected \cite{Moosa:2017yvt}.
\\Recently, in the framework of AdS/CFT two proposals have been introduced to calculate complexity in the gravity side: the Complexity=Volume (CV) \cite{Susskind:2014rva,Stanford:2014jda,Alishahiha:2015rta,Susskind:2018pmk} and the Complexity=Action (CA) \cite{Brown:2015bva,Brown:2015lvg,Susskind:2018pmk} proposals. According to the CA proposal, the holographic complexity $\mathcal{C}$ for a boundary state on a time slice $\Sigma$, is given by the on-shell gravitational action on a region of spacetime called Wheeler-De Witt (WDW) patch, as follows 
\bea
\mathcal{C}(\Sigma) = \frac{I_{WDW}}{\pi \hbar}.
\eea
The WDW patch is defined as the domain of dependence of a Cauchy slice in the bulk which asymptotically approaches the time slice $\Sigma$ on the boundary. In the following, we consider eternal black holes for which the WDW is defined as the domain of dependence of a Cauchy surface started from the right boundary at time $t_R$ and ended on the left boundary at $t=t_L$  (see the left panel of figure \ref{fig:C0}). Then the holographic complexity is associated to the quantum complexity of a state in the dual CFT at time $t=t_L+t_R$.
\\On the other hand, it is well known that complexity is a UV divergent quantity and the structure of its divergences has been studied extensively in the context of both QFT \cite{Jefferson:2017sdb,Chapman:2017rqy,Hackl:2018ptj,Chapman:2018hou,Guo:2018kzl} and holography \cite{Carmi:2016wjl,Reynolds:2016rvl}. In other words, to make a meaningful definition of complexity, one has to regularize it. In QFT one might regularize complexity by putting the QFT on a lattice. In contrast, in the context of holography, two regularization methods were suggested in ref. \cite{Carmi:2016wjl}. In the first regularization shown in the left panel of figure \ref{fig:C0}, the WDW patch is cut at $r=\delta$. Therefore, the WDW patch has two timelike boundaries on the left and right sides. In the second regularization shown in the right panel of figure \ref{fig:C0}, the spacetime is cut at $r=\delta$ and the null boundaries of the WDW patch start from $r=\delta$. 
In ref. \cite{Carmi:2016wjl} it was shown that the structure of the UV divergences are the same in both regularizations, though their coefficients are not equal. On the other hand, in ref. \cite{Reynolds:2016rvl}, the null counterterm $I_{\rm ct}^{(0)}$ (see eq. \eqref{I-ct-0}) was considered, and shown that it removes the ambiguities of the null vectors and at the same time cancels the most divergent term in eq. \eqref{I-0}, however, the coefficients are not equal again.   
\\The first aim of the paper is to show that the two regularization methods are equivalent. To do so, we notice that in the first regularization the WDW patch has two extra timelike boundaries in contrast to that of the second regularization. Indeed, these timelike boundaries are pieces of the boundaries of spacetime. Therefore, one might write some types of counterterms on the timelike boundaries the of WDW patch, which are similar to those applied in holographic renormalization \cite{Balasubramanian:1999re,deHaro:2000vlm,Skenderis:2002wp}. In section \ref{sec: Different Regularizations}, we show that adding these counterterms will resolve the issue of the inequality of the coefficients in the two regularizations.
\\The second aim of the paper is to extract the finite parts of holographic complexity in the CA proposal. Recall that in the CA proposal, the holographic complexity is the gravitational on-shell action evaluated inside the WDW patch. Therefore, it is worth mentioning that in the context of black hole thermodynamics, there are two different approaches for extracting the finite part of the gravitational on-shell action: 1) background subtraction: in this approach one first find a proper background whose asymptotic behavior is the same as that of the black hole. Next, one subtracts the on-shell action of the background from that of the black hole \cite{Gibbons:1976ue,Hawking:1995fd}. Since the asymptotic behaviors of the background and the black hole are the same, the UV divergences of the on-shell action of the black hole are removed, and one finds finite free energy for the black hole. 2) counterterms: in this approach, one adds counterterms applied in holographic renormalization \cite{Balasubramanian:1999re,deHaro:2000vlm,Skenderis:2002wp} to the on-shell action of the black hole, and hence the total action is given by \cite{Emparan:1999pm}
\bea
I_{\rm tot} = I_{\rm bulk} + I_{\rm GHY} + I_{\rm ct}^{\rm HR},
\eea 
where $I_{\rm bulk}$ and $I_{\rm GHY}$ are the bulk action and Gibbons-Hawking-York (GHY) action, respectively (See eq. \eqref{I-bulk} and \eqref{I-GH}). Moreover, $I_{\rm ct}^{\rm HR}$ are the holographic renormalization counterterms given in eq. \eqref{I-ct-HR} which are added on the asymptotic boundary of spacetime. \\Motivated by these approaches, one might apply two different methods to make holographic complexity finite: 1) one might subtract the holographic complexity of the vacuum state, e.g. an AdS spacetime, from that of the given state, e.g. a black hole solution. In this manner, the UV divergences of the state are removed by those of the vacuum state. Actually, this approach is applied in the definition of the complexity of formation in holography \cite{Chapman:2016hwi}, which is a measure of how much hard it is to construct a given excited sate form the vacuum state. For example, for a two-sided black hole the complexity of formation $\Delta \mathcal{C}$ is defined as follows \cite{Chapman:2016hwi}
\bea
\Delta \mathcal{C} = \mathcal{C}_{\rm BH} - 2 \mathcal{C}_{\rm AdS},
\label{compleity-formation}
\eea 
where $\mathcal{C}_{\rm BH} $ and $\mathcal{C}_{\rm AdS} $ are the holographic complexity in the CA proposal for the black hole and the vacuum state, i.e. AdS spacetime, respectively.
\footnote{Since two-sided black holes have two boundaries, one has to subtract the holographic complexity of two copies of AdS in eq. \eqref{compleity-formation}.}
It is shown that at high temperatures and when the spacetime dimension in gravity is higher than three, one has $\Delta \mathcal{C}  \propto S$, where $S$ is the thermal entropy of the corresponding black hole \cite{Chapman:2016hwi}.
2) One might be able to introduce some kind of counterterms on the boundaries of the WDW patch which are able to remove the UV divergences of holographic complexity. In the second part of the paper, we seek these type of counterterms in Einstein gravity. It should be mentioned that, the first attempt in this regard has been made in ref. \cite{Kim:2017lrw}, and the following counterterms were obtained by minimal subtraction 
\bea
I_{\rm ct}=\frac{1}{G_N} \int_{\mathcal{J}} d^{d-1}x \sqrt{h} \sum_{n=0}^{\left[ \frac{d-1}{2}\right]} L^{2n} F_A^{(2n)} \left(d,R_{\mu \nu}, g_{\mu \nu},h_{ij}, K_{ij}\right),
\label{I-ct-Yang}
\eea
here the integral is performed on the null-null joint points $\mathcal{J}$ located on the cutoff surface at $r=\delta$ (see the right panel of figure \ref{fig:C0}). Moreover, $g_{\mu \nu}$ is the induced metric on the $r=\delta$ boundary of spacetime and $R_{\mu \nu}$ is the Ricci tensor made out of $g_{\mu \nu}$. $h_{ij}$ and $K_{ij}$ are the induced metric and the extrinsic curvature tensor of the joint points, respectively. Furthermore, $F_A^{(2n)}$ is a function of an invariant combinations of $\lbrace R_{\mu \nu}, g_{\mu \nu},h_{ij},k_{ij} \rbrace$ and is of mass dimension $2n$. 
In the following, we want to introduce new counterterms which are written on the boundaries of the WDW patch which are codimension-one null surfaces. Moreover, we want to write counterterms which are covariant and do not change the equations of motion.
\\The organization of the paper is as follows: in section \ref{sec: Setup}, we fix our notations. In section \ref{sec: Different Regularizations}, we consider two different methods for regularizing holographic complexity (see figure \ref{fig:C0}), and argue they are completely equivalent. In other words, we show the structure of the UV divergences of the on-shell action as well as their coefficients are the same. In section \ref{sec: General Form of  Null Counterterms}, we discuss on the general form of the counterterms on null boundaries of the WDW patch in an asymptotically AdS spacetime. These counterterms are able to remove all the UV divergences of holographic complexity. In section \ref{sec: Holographic Complexity}, we calculate the null counterterms for an AdS-Schwarzschild black hole in Einstein gravity. In section \ref{sec: Complexity of Formation}, we compute the complexity of formation for an AdS-Schwarzschild black hole for $d=2,3,4$, and show that the aforementioned null counterterms do not change the complexity of formation studied in ref. \cite{Chapman:2016hwi}. In section \ref{sec: New Counterterm on the Singularity}, we introduce another counterterm on the future singularity of the black hole which is able to remove an IR logarithmic divergent term, i.e. $\log r_{\rm max}$, created by the null counterterms for odd $d$. In section \ref{sec: Growth Rate of Complexity}, we study the effect of the null counterterms on the growth rate of holographic complexity. At the end, in section \ref{sec: Discussion}, we conclude and discuss about charged black holes. 
\section{Setup}
\label{sec: Setup}
In this section, we fix our notations. For simplicity, in the following we restrict ourselves to an Einstein gravity in $d+1$ dimensions whose action is written as follows 
\bea\label{I-bulk}
I_{\rm bulk}=\frac{1}{16\pi G_N}\int d^{d+1}x\sqrt{-g}\left(R-2 \Lambda
\right)\,,
\eea
where $G_N$ is the Newton's constant and $\Lambda= -\frac{d(d-1)}{L^2}$ is the cosmological constant in which $L$ is the AdS radius of curvature. This action has an AdS-Schwarzschild black hole solution whose metric may be parametrized by
\bea
ds^2 =\frac{L^2}{r^2} \left(-f(r) dt^2+ \frac{1}{f(r)} dr^2 +L^2 d\Omega^2_{d-1}\right),\;\;\;\;\;
\;\;\;
f(r)=1+ \frac{r^2}{L^2}-\frac{r^d}{r_0^d},
\label{metric-AdS-Schwarzschild}
\eea
where $d\Omega^2_{d-1}$ is the metric of a unit $d-1$ dimensional sphere, and $r_0$ is 
related to the radius of the horizon $r_h$ via
\bea
r_0^{-d}= {r_h^{-d}}\left(1+\frac{r_h^2}{L^2}\right)\,.
\label{horizon}
\eea
Moreover, we define tortoise coordinate $r^*(r)$ as follows
\bea
r^*(r_2)-r^*(r_1)= -\int_{r_1}^{r_2} \frac{dr}{f(r)}.
\label{tortoise}
\eea 
To calculate the holographic complexity, one needs to compute on-shell gravitational action on the WDW patch. The action is composed of different parts as follows \cite{Lehner:2016vdi} 
\bea\label{I-0}
I = I_{\rm bulk}+ I_{\rm GHY}+ I_{\rm joint}+ I_{\rm ct}^{(0)}.
\eea
In the following, we will introduce each part. In general, the WDW patch has timelike, spacelike, and null boundaries, which are codimension-one hypersurfaces and we show them by $\mathcal{T}, \mathcal{S}, \mathcal{N}$, respectively. The extrinsic curvature of the corresponding boundaries are denoted by $K_t, K_s$ and $K_n$, and one has to include a Gibbons-Hawking-York (GHY) term  \cite{York:1972sj,Gibbons:1976ue} for each boundary
\bea
I_{\rm GHY} = \frac{1}{8\pi G_N}
\int_{\mathcal{T}} K_t\; d\Sigma_t \pm\frac{1}{8\pi G_N} \int_{\mathcal{S}} K_s\; d\Sigma_s
\pm \frac{1}{8\pi G_N} \int_{\mathcal{N}} K_n\; dS
d\lambda\; .
\label{I-GH}
\eea 
In the GHY term for null surfaces $\lambda$, is the coordinate on the null boundaries. In the following, we choose $\lambda$ to be affine, hence the GHY action will be zero on null boundaries. There are also some joint points where two boundaries intersect each other. The joints shown by $\mathcal{J}$ are codimension-two hypersurfaces and their action is given in terms of the function $a$ which is given by the logarithm of the inner product of the normal vectors to the corresponding boundaries \cite{Hayward:1993my,Brill:1994mb}
\bea
I_{\rm joint} = \pm\frac{1}{8\pi G_N} \int_{\mathcal{J}} a\; dS.
\label{I-Joint}
\eea
The sign of different terms in action, depend on the relative position of the boundaries and the bulk region of interest (see \cite{Lehner:2016vdi} for more details). 
\\Moreover, it was shown that null boundaries of spacetime contribute to the action \cite{Parattu:2015gga,Chakraborty:2016yna,Lehner:2016vdi,Chakraborty:2018dvi}. In particular, in ref. \cite{Lehner:2016vdi} it was observed that there is an ambiguity in the normalization of normal vectors to null boundaries, and one has to introduce the following counterterm,
\bea
I^{(0)}_{\rm ct} = \frac{1}{8 \pi G_N} \int_{\mathcal{N}} d \lambda d^{d-1} \Sigma \sqrt{\gamma} \Theta \ln  | \tilde{L} \Theta | ,
\label{I-ct-0}
\eea 
on the null boundaries to remove the ambiguities. Here, $\gamma$ is the determinant of the induced metric and the quantity $\Theta$ is the expansion of the null generators which is defined as follows
\bea
\Theta=\frac{1}{\sqrt{\gamma}}\frac{\partial\sqrt{\gamma}}{\partial\lambda},
\label{Theta}
\eea
where the parameter $\tilde{L}$ is an undetermined length scale, which can have any value. In the dual QFT, the ambiguity in the definition of $\tilde{L}$ is related to the freedom in choosing the reference state \cite{Jefferson:2017sdb,Chapman:2017rqy}. In other words, one can write $\tilde{L} = M L$, where $M$ is the scale of the reference state and $L$ is the AdS radius of curvature  \cite{Jefferson:2017sdb,Chapman:2017rqy}. In the following, we have fixed  $\tilde{L}$ to be $\tilde{L}=\frac{L}{d-1}$ for convenience.
\footnote{In holography, there are other choices for this length scale, for example in refs. \cite{Reynolds:2016rvl,Kim:2017lrw} it is set to $\tilde{L}=L$.}
It should be stressed that this choice will remove a UV divergent term of order $\mathcal{O}(\delta^{-d+1})$ in eq. \eqref{I-ct-0} and hence in the holographic complexity. As we will see, the counterterm given in eq. \eqref{I-ct-0} together with other counterterms play a crucial role in order to get the desired results.
\\Moreover, in the following, we set the time on the left and right boundaries as $t_L=t_R= \frac{t}{2}$ to have a symmetric WDW patch shown in figure \ref{fig:C0}. Moreover, we calculate holographic complexity at times $t>t_c$, when the past light sheets from the left and right boundaries of spacetime do not touch the past singularity. In tis case, it is straightforward to show that the critical time $t_c$ is given by $t_c=2 \left( r^*(\delta) - r^*(r_{\rm max}) \right)$ \cite{Carmi:2017jqz}. 
\section{Different regularizations}
\label{sec: Different Regularizations}
\begin{figure}
	\begin{center}
		\includegraphics[scale=1]{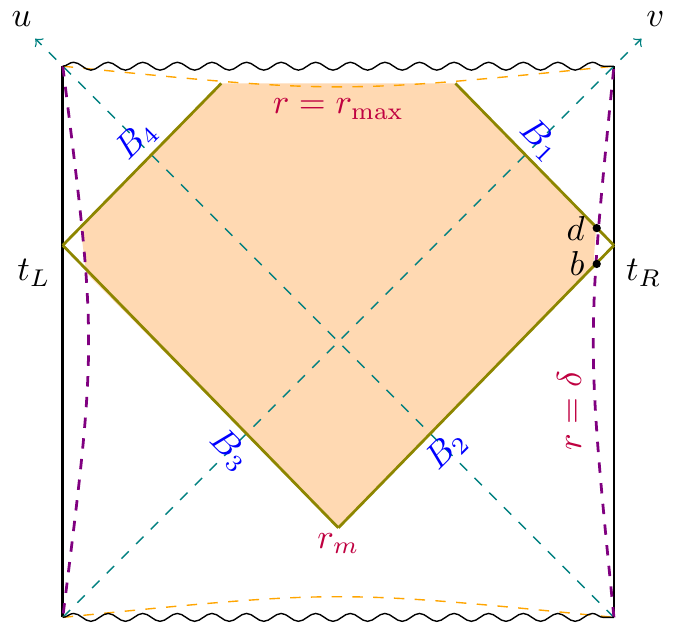}
		\includegraphics[scale=1]{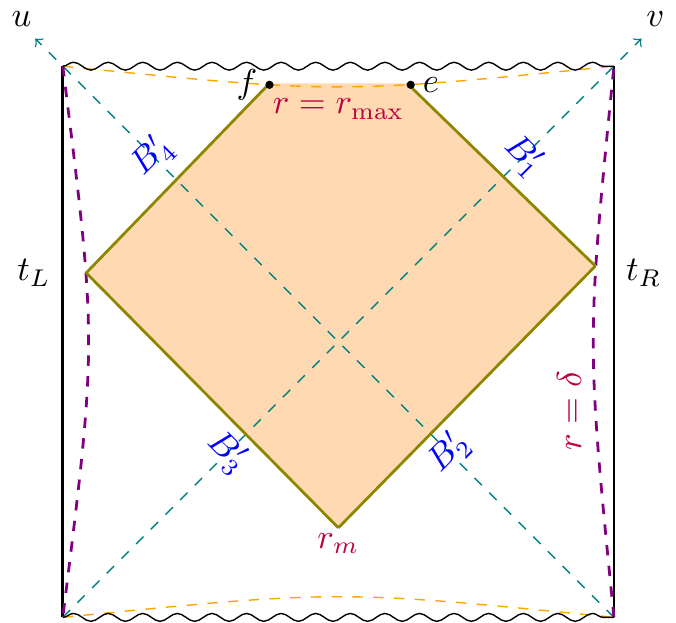}
	\end{center}
	\caption{WDW patches for an eternal two-sided black hole in two different regularizations: Left) the first regularization in which we cut 
		the WDW patch at the radius $r=\delta$. Right) the second regularization in which we cut the spacetime 
		at the radius $r=\delta$.}
	\label{fig:C0}
\end{figure}

In this section, we will study the UV divergences of the holographic complexity of an eternal two-sided AdS-Schwarzschild black hole in Einstein gravity by applying the CA proposal for two different  regularizations shown in figure \ref{fig:C0}.  In the first regularization, we will cut the WDW patch at the radius $r=\delta$ (See the left panel of figure \ref{fig:C0}), while in the second regularization, we will cut the spacetime at $r=\delta$ (See the right panel of figure \ref{fig:C0}). The aim is to verify that the two different regularizations are equivalent, in the sense that holographic complexity have the same UV divergence structure with the same coefficients in both of them. 
\\It should be pointed out that, these regularizations have already been studied for global $AdS_{d+1}$ spacetimes in \cite{Carmi:2016wjl,Reynolds:2016rvl}. Indeed, it has been shown that the structure of the UV divergences in the two regularizations are the same but their coefficients are different. Looking at the two WDW patches in figure \ref{fig:C0}, one observes that in the first regularization, the WDW patch have two extra timelike boundaries at $r=\delta$ (one on the left hand side and the other on the right hand side of the WDW patch). Here we want to show that by adding some types of counterterms (See eq. \eqref{I-ct-HR-1}) similar to those applied in holographic renormalization, and the corresponding GHY term for these two timelike boundaries, not only the structure of the UV divergences, but also their coefficients become exactly the same. In the following, we consider the AdS-Schwarzschild solution \eqref{metric-AdS-Schwarzschild}. It is evident that the UV divergence structure of holographic complexity comes from the asymptotic behavior of the solution, and if one considers a pure AdS spacetime instead of \eqref{metric-AdS-Schwarzschild}, one should obtain the same result.  
\\Having fixed our notations, now we calculate the holographic complexity for the geometry given in eq. \eqref{metric-AdS-Schwarzschild} using the two regularizations. To proceed let us first write the equations for the null boundaries $B_i$ of the corresponding WDW patches. It is straightforward to check that for the first regularization, one has
\bea
&&B_1:\,\,t=t_R+r^*(0)-r^*(r),\;\;\;\;\;\;\;\;\;\;B_2:\,\,t=t_R-r^*(0)+r^*(r),
\cr &&\cr
&&B_3:\,\,t=-t_L+r^*(0)-r^*(r),\;\;\;\;\;\;\;\;B_4:\,\,t=-t_L-r^*(0)+r^*(r),
\label{null-bdy-reg-1}
\eea
while for the second regularization, one has
\bea
&&B'_1:\,\,t=t_R+r^*(\delta)-r^*(r),\;\;\;\;\;\;\;\;\;\;B'_2:\,\,t=t_R-r^*(\delta)+r^*(r),
\cr &&\cr
&&B'_3:\,\,t=-t_L+r^*(\delta)-r^*(r),\;\;\;\;\;\;\;\;B'_4:\,\,t=-t_L-r^*(\delta)+r^*(r).
\label{null-bdy-reg-2}
\eea
With this notation, and using the fact that for this metric, one has
\be
\sqrt{-g}(R-2\Lambda)=-2d\frac{L^{2(d-1)}}{r^{d+1}},
\ee
the contribution of the bulk action in the first and second regularizations are given by
\bea\label{B0}
I^{\rm reg.1}_{\rm bulk} = -\frac{V_{d-1}L^{2(d-1)}d}{4\pi G_N}\bigg( \!\!&\!\! 2 \!\!&\!\! \int_\delta^{r_{\rm Max}} 
\frac{dr}{r^{d+1}}
(r^*(0)-r^*(r))  
\cr && \cr
\;\;\;\;\;\;\;\;\;\;\;\; &\!\!+&\!\! \int_{r_m}^{r_{\rm Max}} \frac{dr}{r^{d+1}} 
\left( \frac{t}{2}-r^*(0)+r^*(r) \right) \bigg),
\cr &&\cr
I^{\rm reg.2}_{\rm bulk}=-\frac{V_{d-1}L^{2(d-1)}d}{4\pi G_N}\bigg(\!\!&\!\! 2 \!\!&\!\! \int_\delta^{r_{\rm Max}} 
\frac{dr}{r^{d+1}}
(r^*(\delta)-r^*(r)) 
\nonumber\\
&+& \int_{r_m}^{r_{\rm Max}} \frac{dr}{r^{d+1}} 
\left( \frac{t}{2}-r^*(\delta)+r^*(r) \right) \bigg).
\eea
Here $V_{d-1}$ is the volume of a unit $d-1$ dimensional sphere, i.e. $V_{d-1}= \frac{2 \pi^{\frac{d}{2}}}{\Gamma \left( \frac{d}{2} \right)}$. Since the second terms in the above expressions are finite, and we are only interested in comparing the divergent structure of the holographic complexity, we just need to consider the first terms in these expressions. Indeed, by adding and subtracting $r^*(\delta)$ to the divergent part of the first regularization, one finds
\bea
(I_{\rm bulk}^{\rm reg. 1})|_{\rm div} &\!= &\!-\frac{V_{d-1}L^{2(d-1)}}{2\pi G_N}
\frac{r^*(0)-r^*(\delta)}{\delta^{d}}
 -\frac{V_{d-1}L^{2(d-1)}d}{2\pi G_N}\int_\delta^{r_{\rm Max}} 
\frac{dr}{r^{d+1}}
(r^*(\delta)-r^*(r))\cr &&\cr
 \!&\!\!=\!\!&\!-\frac{V_{d-1}L^{2(d-1)}}{2\pi G_N}
\frac{ ( r^*(0)-r^*(\delta))}{\delta^{d}}
 +(I_{\rm bulk}^{\rm reg.2})|_{\rm div} \; .
\label{I-bulk-1-2}
 \eea
Therefore, as far as the bulk term is concerned, there is a difference between the UV divergences  of the two regularizations.
\\Now we consider the contribution of the joint points to the action. In the first regularization, there are two timelike-null joints $b$ and $d$ on the right timelike boundary (see the left panel of figure \eqref{fig:C}) whose actions are as follows
\bea
I_{\rm joints}^{\rm reg.1}=- \frac{1}{8 \pi G_N} \int_{b} d^{d-1} \sigma \sqrt{h} \log |k_1 . s |- \frac{1}{8 \pi G_N} \int_{d} d^{d-1} \sigma \sqrt{h} \log |k_2 . s |,
\eea 
where $k_1$ and $k_2$ are the normal vectors to the null surfaces $B_1$ and $B_2$ in eq. \eqref{null-bdy-reg-1}, and are given by
\bea
k_1= \alpha \left(dt - \frac{dr}{f(r)} \right), \;\;\;\;\;\;\;\;\;\;\;\;  k_2= \beta \left(dt + \frac{dr}{f(r)} \right).
\eea 
we choose the normalization of the normal vectors such that $k_i .\hat{t}=c$, in which $\hat{t}=\partial_t$ and $c$ is a positive constant \cite{Lehner:2016vdi}. On the other hand, $s$ is the spacelike outward-directed normal vector to the timelike boundary at $r=\delta$,
\bea
s = -\frac{L}{\delta \sqrt{f(\delta)}} dr.
\eea 
It is straightforward to show that in the first regularization we have
\bea
I_{\rm joints}^{\rm reg.1}= - \frac{V_{d-1} L^{2(d-1)}}{8 \pi G_N} \frac{1}{\delta^{d-1}} \log \frac{\alpha \beta}{L^2 f(\delta)}.
\eea 
On the other hand, in the second regularization there is a null-null joint $e$, on the right hand side of the WDW patch, whose action is given by
\bea
I_{\rm joints}^{\rm reg.2}= -\frac{1}{8 \pi G_N} \int_{e} d^{d-1} \sigma \sqrt{h} \log \frac{| k^\prime_1 . k^\prime_2|}{2},
\eea 
here $k^\prime_1$ and $k^\prime_2$ are the normal vectors to the null boundaries $B^\prime_1$ and $B^\prime_2$ in eq. \eqref{null-bdy-reg-2}, respectively. It is evident that $k^\prime_1= k_1$ and $k^\prime_2= k_2$. Therefore, we have
\bea
I_{\rm joints}^{\rm reg.2} &=&  - \frac{V_{d-1} L^{2(d-1)}}{8 \pi G_N} \frac{1}{\delta^{d-1}} \log \frac{\alpha \beta}{L^2 f(\delta)}
\cr && \cr
&=& I_{\rm joints}^{\rm reg.1}
\label{I-joint-12}
\eea 
Now we consider the counterterm $I_{\rm ct}^{(0)}$. One can find the the expansions $\Theta_i$ and affine parameters $\lambda_i$ of the null boundaries in the first regularization as follows
\bea
\Theta_1 &=& \frac{\alpha (d-1) r}{L^2},  \;\;\;\;\;\;\;\;\;\;\;\;\;\;\;\;\;\;\;\;\;\;\;\;  \Theta_2 = -\frac{\beta (d-1) r}{L^2},
\cr && \cr
\lambda_1 &=&  \frac{L^2}{\alpha r}, \;\;\;\;\;\;\;\;\;\;\;\;\;\;\;\;\;\;\;\;\;\;\;\;\;\;\;\;\;\;\;\;\;\;\;  \lambda_2 = - \frac{L^2}{\beta r}.
\label{null-vectors}
\eea 
Moreover, since in the two regularizations the normal vectors, and hence their null expansions $\Theta_i$ are the same, one can conclude that the counterterms $I_{\rm ct}^{(0)}$ are equal in these regularizations,
\bea
I_{\rm ct}^{(0), \rm reg.1} = I_{\rm ct}^{(0),\,  \rm reg2}\;.
\label{I-ct0-12}
\eea 
Of course, this is not the whole story. Indeed, as said before, in the first regularization the WDW patch has two extra timelike boundaries at $r=\delta$, in comparison to the WDW patch in the second regularization (See figure \eqref{fig:C0}). Therefore, in the first regularization, one should consider the corresponding action for each of these timelike boundaries. Naturally one can write a GHY term, eq. \eqref{I-GH-2}, on each of them. On the other hand, from holographic renormalization one can write the following counterterms on the whole boundary of spacetime at $r=\delta$ \cite{Balasubramanian:1999re,deHaro:2000vlm,Skenderis:2002wp,Emparan:1999pm}.
\footnote{
We should point out that in our notation the Ricci tensor and Ricci scalar have an extra minus sign with respect to those of \cite{Balasubramanian:1999re,deHaro:2000vlm}.
}
\bea
I_{\rm ct}^{\rm HR} = -\frac{1}{16 \pi G} \int_{r=\delta} d^{d-1}x dt\;\sqrt{-h}\;\left (\frac{2(d-1)}{L}
+\frac{L}{(d-2)}{\cal R} + a_{(d)} \log \delta + \cdots\right),
\label{I-ct-HR}
\eea
where $h$ is the determinant of the induced metric on the $r=\delta$ surface and ${\cal R}$ is the corresponding Ricci scalar. Moreover, the logarithmic counterterm exists for even d, and its coefficient $a_{d}$ is related to the conformal anomaly of the dual CFT \cite{deHaro:2000vlm,Henningson:1998gx}. One should note that the timlelike boundary of the WDW patch in the first regularization is a finite piece of the whole boundary of the spacetime. Therefore, inspired by holographic renormalization, one might consider the following counterterms on the timelike boundaries of the WDW patch in the left panel of figure \ref{fig:C0}
\bea
I_{\rm ct}^{\mathcal{T}} = -\frac{1}{16 \pi G} \int_{r=\delta} d^{d-1}x dt\;\sqrt{-h}\;\left (\frac{2(d-1)}{L}
+\frac{L}{(d-2)}{\cal R} + \cdots\right).
\label{I-ct-HR-1}
\eea
Here since we do not have any logarithmic divergent terms in the action $I$ (See eq. \eqref{I-0}), we do not apply the logarithmic counterterms in eq. \eqref{I-ct-HR}. As we will see in the following, it is crucial to include the above timelike counterterms to show that the coefficients of the UV divergences of the on-shell action in the two regularizations are exactly the same.
\\Now we calculate the divergent parts of the GHY term \eqref{I-GH-2} and timelike counterterms \eqref{I-ct-HR-1} on $r=\delta$ surfaces, respectively. The GHY term is given by
\bea
I_{\rm GHY} = 2\times\frac{1}{8\pi G} \int_{r=\delta} d^{d-1}x dt\;\sqrt{-h} K,
\label{I-GH-2}
\eea
where the factor of two is included to account for the contributions of the left and right timelike boundaries at $r=\delta$, 
$h$ is the determinant of the induced metric on the timelike boundary, and $K$ is its extrinsic curvature. One can write
\bea
\sqrt{-h} |_{r=\delta}&=& \frac{L^{2d-1}}{\delta^d} \left(1+ \frac{\delta^2}{2 L^2} + \cdots \right) ,
\nonumber\\
K&=&\left( \frac{2df(r)-r f'(r)}{2 L \sqrt{f(r)}} \right)|_{r=\delta}
\nonumber\\
&=& 2d+\frac{2(d-1) \delta^2}{L^2}- \frac{d \delta^d}{r_0^d}.
\eea
Note that in eq. \eqref{I-GH-2} the integral on the time coordinate is taken on an interval from the past to the future null boundaries which is given by
\bea
\Delta t=  t_d - t_b = 2 \left(r^*(0) - r^*(\delta) \right).
\label{Delta-t}
\eea 
Then it is straightforward to compute the divergent part of the GHY term,
\bea
I_{\rm GHY}|_{\rm div} =\frac{V_{d-1}L^{2(d-1)}}{2\pi G}\left(
\frac{d}{\delta^d}+\frac{(d-1)}{L^2\delta^{d-2}}\right)(r^*(0)-r^*(\delta)).
\label{I-GH-timelike}
\eea
Now we consider the timelike counterterms \eqref{I-ct-HR-1}. By applying
\bea
\mathcal{R} |_{r=\delta}= \frac{(d-1)(d-2) \delta^2}{L^4} ,
\label{Ricci-scalar-tensor-timelike-bdy}
\eea 
one can easily write
\bea 
I_{\rm ct}^{\mathcal{T}} =- \frac{(d-1) V_{d-1}L^{2(d-1)}}{2\pi G}
\left(\frac{1}{\delta^d}+\frac{1}{L^2\delta^{d-2}}+\cdots\right) \; (r^*(0)-r^*(\delta)),
\label{I-ct-HR-2}
\eea
here a factor of two is included to consider the contributions of the left and right timelike boundaries at $r=\delta$. By adding eq. \eqref{I-GH-timelike} to eq. \eqref{I-ct-HR-2}, one has
\bea
I_{\rm GHY}+ I_{\rm ct}^{\mathcal{T}} = \frac{V_{d-1} L^{2(d-1)}}{2 \pi G_N} \frac{( r^*(0)-r^*(\delta))}{\delta^{d}} .
\label{I-GH+HR}
\eea 
It is then evident that these divergent terms cancel those coming from the bulk term in eq. \eqref{I-bulk-1-2}, and leads to
\footnote{Note that as one goes to higher dimensions more counterterms are needed, though the general structure is the same as  what is demonstrated.} 
\bea
(I_{\rm bulk}^{\rm reg.1}+ I_{\rm GHY} + I_{\rm ct}^{\mathcal{T}})|_{\rm div} = I_{\rm bulk}^{\rm reg. 2}|_{\rm div} \; .
\label{I-bulk12+GH-HR}
\eea 
Therefore, from eq. \eqref{I-joint-12}, \eqref{I-ct0-12} and eq. \eqref{I-bulk12+GH-HR}, one can conclude that the divergent parts of the total action in the two regularizations are equal to each other,
\be
I^{\rm reg. 1}|_{\rm div}=I^{\rm reg.2} |_{\rm div}.
\label{I-reg1=I-reg2}
\ee
Therefore, in both regularizations the structure and coefficients of the UV divergences of  holographic complexity are exactly the same, provided that one takes into account all surface terms including the counterterms inspired by holographic renormalization, i.e. eq. \eqref{I-ct-HR-1}. We should stress that although we have derived eq. \eqref{I-reg1=I-reg2} for the specific value of $\tilde{L} = \frac{L}{d-1}$, it holds for any value of $\tilde{L}$.
\\To the best of our knowledge, the holographic renormalization counterterms have never been considered before in the literature of holographic complexity. Moreover, we used them on a small time interval of the AdS boundary, which is one of the boundaries of the WDW patch. Therefore, it seems that in the calculation of the on-shell action in any region of spacetime, it is necessary to consider the role of counterterms on all boundaries of that region.
\footnote{See also \cite{Frassino:2019fgr} for a discussion on the role of counterterms in the thermodynamics and holographic complexity of  exotic BTZ black holes in gravitational Chern-Simons theory.}
Since, the calculation of holographic complexity in the second regularization is easier, in the rest of the paper, we apply it.
\section{General form of  null counterterms}
\label{sec: General Form of  Null Counterterms}
\begin{figure}
\begin{center}
		\includegraphics[scale=1]{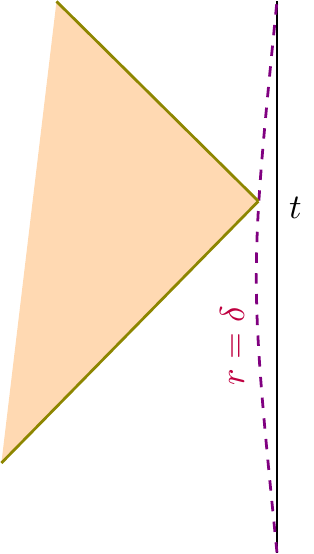}
\end{center}
	\caption{One half of the WDW patch near the right boundary of spacetime.}
	\label{fig:C}
\end{figure}

In the previous section, we discussed the important role of counterterms in the equivalence of the regularizations. The aim of this section is to explore new types of counterterms on null boundaries of the WDW patch which are able to remove all the UV divergences of holographic complexity. We should also emphasize that using the minimal subtraction scheme, certain counterterms have been introduced in \cite{Kim:2017lrw}, which could make the on-shell action finite. However, those counterterms are written on joint points of the WDW, and are not on the codimension-one boundaries of the WDW. In what follows, we would like to revisit the procedure and find new types of counterterms which are: covariant, written on the null boundaries, and do not change the equations of motion.
 \hspace{3mm}
\\Our strategy is to first extract the UV divergent terms of the on-shell action, and then to rewrite them in terms of the intrinsic and extrinsic properties of the null boundaries. Next, we apply the minimal subtraction scheme and introduce the appropriate counterterms. 
In \cite{Carmi:2016wjl,Reynolds:2016rvl} these divergent terms have been calculated for an asymptotically $AdS_{d+1}$ spacetime in Fefferman-Graham coordinates. To study the general form of the counterterms, we consider an asymptotically AdS geometry whose metric in the Fefferman-Graham coordinates is as follows
\be
ds^2=G_{\mu \nu} dx^\mu dx^\nu = \frac{L^2}{z^2}(dz^2+g_{ij}(z,x)dx^idx^j),
\ee
where \cite{deHaro:2000vlm,Skenderis:2002wp}
\bea
\label{g-0-ij}
g_{ij}(z,x) &=& g_{ij}^{(0)}(x)+z^2 g_{ij}^{(1)}(x)+\cdots,
\\
g_{ij}^{(1)} &=&\frac{1}{(d-2)}\left(R_{ij}-\frac{g_{ij}^{(0)}}{2(d-1)}R\right).
\label{g-1-ij}
\eea
Here $z$ is the radial coordinate and the boundary is located at $r=\delta$. Moreover, $R_{ij}$ and $R$ are  Ricci tensor and Ricci scalar constructed  out of  $g^{(0)}_{ij}$. Since we are interested in computing  the on-shell action on a subspace ({\it e.g.} WDW
patch) that could contain several null, spacelike, timelike boundaries as well as 
their intersections, we will have to  consider several codimension-one and codimension-two
boundaries. It is then crucial to write the final action in a covariant way to make sure that 
the new counterterms will not alter the variational principle. To proceed it is useful to decompose the coordinates $x^i$ into $ t$ and $\sigma^a$
for $a=1\cdots d-1$. Assuming $g_{tt}^{(0)}=-1$ and $g_{ta}^{(0)}=0$, one has
\bea
R_{tt}&=&-\frac{1}{4}{\rm Tr}\left[\left((g^{(0)})^{-1}\partial_tg^{(0)}\right)^2\right]+\frac{1}{2}
{\rm Tr}[(g^{(0)})^{-1}\partial^2_tg^{(0)}],\\ &&\cr
R_{ab}&=&\frac{1}{4}\partial_tg^{(0)}_{ac}g^{(0)cd}\partial_tg^{(0)}_{bd}
+\frac{1}{4}\partial_tg^{(0)}_{bc}g^{(0)cd}\partial_tg^{(0)}_{ad}
-\frac{1}{2}\partial_t^2g^{(0)}_{ab}-\frac{1}{4}\partial_tg^{(0)}_{ab}g^{(0)cd}\partial_tg^{(0)}_{cd}
+{\cal R}_{ab},\nonumber
\eea
where ${\cal R}_{ab}$ is the Ricci tensor of the joint points where two null boundaries intersect. On the joint points the coordinates are given by $\sigma^a$. Moreover, one gets
\bea\label{gg}
g^{(1)} _{ab}&=&\frac{1}{(d-2)}\bigg(R_{ab}-\frac{g_{ab}^{(0)}}{2(d-1)}(-R_{tt}+ g^{(0)ab}
R_{ab})\bigg),
\cr &&\cr
g_{tt}^{(1)}&=&\frac{1}{(d-2)}\bigg(R_{tt}+\frac{1}{2(d-1)}(-R_{tt}+ g^{(0)ab}R_{ab})\bigg),
\eea
such that
\bea
g^{(0)ab}g^{(1)} _{ab}&=&\frac{1}{2(d-2)}( g^{(0)ab}R_{ab}+R_{tt})\\
&=&\frac{1}{2(d-2)}\bigg(\frac{1}{4}{\rm Tr}\left[\left((g^{(0)})^{-1}\partial_tg^{(0)}\right)^2\right]-
\frac{1}{4} \left(Tr[(g^{(0)})^{-1}\partial_tg^{(0)}]\right)^2+\cal R\bigg),
\eea
where ${\cal R}=g^{(0)ab}{\cal R}_{ab}$ is the Ricci scalar of the joint point. In what follows, it is also useful to expand the determinant of the asymptotic metric
around $t=0$. Indeed, applying eq. \eqref{g-0-ij} one has 
\be
g_{ij}(t,z,\sigma)= h_{ij}(\sigma)+{\partial_t g_{ij}^{(0)}(t,\sigma)}|_{t=0}t+
\frac{1}{2}
{\partial^2_t g_{ij}^{(0)}(t,\sigma)}|_{t=0}t^2+z^2g_{ij}^{(1)}(t=0,\sigma)+\cdots ,
\label{g-ij-expansion}
\ee
where  $h_{ij}(\sigma)=g^{(0)}_{ij}(t=0,\sigma)$. Therefore, one obtains
\bea
\sqrt{\det g_{ij}}\!&\!=\!&\!\sqrt{h}\,\bigg[ 1+\frac{t}{2}{\rm Tr}(h^{-1}\partial_{t}g^{(0)})+\frac{t^2}{4}
\big(
{\rm Tr} (h^{-1}\partial_{t}^2g^{(0)})
\\
&&\;\;\;\;\;\;\;\;\;\;\;\;\;\; -{\rm Tr}(h^{-1}\partial_{t}g^{(0)})^2
+\frac{1}{2}({\rm Tr} (h^{-1}\partial_{t}g^{(0)}))^2\bigg)
+\frac{z^2}{2}{\rm Tr}(h^{-1}g^{(1)})+\cdots \bigg],
\nonumber
\label{g-z-t}
\eea
which can be recast into the following form \cite{Carmi:2016wjl}
\be
\sqrt{\det g_{ij}}=\sqrt{h}\,\bigg(1+t\,q^{(0)}_{1}+t^2\,q^{(0)}_{2}+z^2\,q^{(2)}_{0}+\cdots\bigg),
\ee
with the identifications of
\bea\label{qs}
&&q^{(0)}_{1}=\frac{1}{2}{\rm Tr}(h^{-1}\partial_{t}g^{(0)}),\;\;\;\;\;\;\;\;\;\;\;\;\;\;\;\;\;\;\;\;
q^{(2)}_{0}=\frac{1}{2}{\rm Tr}(h^{-1}g^{(1)}),\cr &&\cr
&&q^{(0)}_{2}=\frac{1}{4}
\left(
{\rm Tr} (h^{-1}\partial_{t}^2g^{(0)})
-{\rm Tr}(h^{-1}\partial_{t}g^{(0)})^2
+\frac{1}{2}({\rm Tr} (h^{-1}\partial_{t}g^{(0)}))^2\right)\,.
\eea
The last ingredient we need to compute the on-shell action in the WDW patch is the {\it extrinsic curvature} along the null boundaries. In our coordinate system, the induced metric on a null boundary $\mathcal{N}$ may be written as follows
\be
ds_{\mathcal{N}}^2=\gamma_{ij}dx^idx^j=\frac{L^2}{z^2}
\left(g_{ta}dtd\sigma^a+g_{ab}d\sigma^a d\sigma^b\right),
\label{induced-metric-null-1}
\ee
with the assumptions that
\be
g_{ta}={\cal O}(z^2),\;\;\;\;\;\;\;\;g_{ab}=h_{ab}+z^2 g_{ab}^{(1)}+\cdots\,.
\ee
In the following, we work with the second regularization and set $t_L=t_R=0$, then near the asymptotic boundary at $z=\delta$, the future $B^\prime_1$ and past $B^\prime_2$ null boundaries (See figure \ref{fig:C}) are given by \cite{Carmi:2016wjl}
\bea
B^\prime_1:  \;\; t&=&t_+(z,\sigma ) = \;\;\; ( z-\delta) +\frac{(z^3-\delta^3)}{6}g^{(1)}_{tt}+\cdots, \;\;\;\;\; \text{for $t \geq 0$},
\nonumber\\
B^\prime_2:  \;\; t&=&t_{-} (z,\sigma) = -(z-\delta)-\frac{(z^3-\delta^3)}{6}g^{(1)}_{tt}+\cdots, \;\;\;\;\; \text{for $t \leq 0$}
\label{null-surfaces-FG}
\eea 
and their normal vectors are given by $k_F= \alpha (dt-dt_{+})$ and $k_P =-\beta (dt - dt_{-} )$. Here $\alpha$ and $\beta$ are constants that appear due to the ambiguity in the normalization of the normal vector to null surface $\mathcal{N}$. Moreover, we need to calculate the affine parameter $\lambda$ of the null surfaces. For future null boundary, the affine parameter to the order that we are interested in here, is given by eq. \eqref{Lambda-affine-future} (see appendix \ref{App-A})
\footnote{
We should point out that in \cite{Kim:2017lrw} a non-affine parameter is used for null boundaries, and hence the authors had to consider the GHY term on null boundaries. Here, since we found the correct affine parameter, the GHY term on null boundaries is zero.
}
\bea
\lambda=\frac{L^2}{\alpha z}\left(1+\frac{z^2}{2}g_{tt}^{(1)}+{\cal O}(z^4)\right).
\eea
In this notation the object we are looking for  may be defined as follows
\be
\Theta^i_{j}=\frac{1}{2}\gamma^{ik}\partial_{\lambda}\gamma_{kj}.
\label{Theta-ab}
\ee
In what follows, we will have to deal with the trace and inner product of the extrinsic curvature tensor $\Theta^i_j$ which are defined as $\Theta=\Theta^i_i$ and $\Theta\cdot \Theta=\Theta^i_j \Theta^j_i$, respectively. We note, however, that  since the metric component $g_{ta}$ starts at order ${\cal O}(z^2)$, the component $\Theta^t_t$ starts at order ${\cal O}(z^5)$. Therefore,
up to the order ${\cal O}(z^3)$ that we are interested in here, it is sufficient to work with $\Theta=\Theta^a_a$ and $\Theta\cdot \Theta=\Theta^a_b \Theta^b_a$. It is then straightforward to compute these objects using the asymptotic behavior of the metric. In particular, for the future null boundary we have 
\be
t=z-\delta+\frac{z^3-\delta^3}{6}g^{(1)}_{tt}+\cdots,
\label{Null-past}
\ee
If one plugs eq. \eqref{Null-past} into eq. \eqref{g-z-t}, one has
\be
g_{cb}=h_{cb}(\sigma)+{\partial_t g^{(0)}_{cb}(t,\sigma)}|_{t=0}(z-\delta)
+\frac{1}{2}{\partial^2_t g^{(0)}_{cb}(t,\sigma)}|_{t=0}(z-\delta)^2+z^2g^{(1)}_{cb}
(t=0,\sigma),
\ee
Next, one can find 
\bea
\Theta^a_{b} &=&\frac{\alpha z}{L^2}\bigg(1-\frac{z}{2}h^{-1}\partial_t g^{(0)}-\frac{1}{2}z(z-
\delta)h^{-1}\partial^2_{t} g^{(0)}
\cr && \cr && \;\;\;\;\;\;\; -z^2h^{-1}g^{(1)}
+\frac{1}{2}z(z-\delta)(h^{-1}\partial_t 
g^{(0)})^2+\frac{z^2}{2}g_{tt}^{(1)}+...\bigg)^a_{b}.
\eea
Then it is straightforward to write
\bea
\Theta=\Theta^a_{a}&=&\frac{\alpha z}{L^2}\bigg((d-1)-\frac{1}{2}z{\rm Tr}(h^{-1}
\partial_t g^{(0)})-\frac{1}{2}z(z-\delta){\rm Tr}(h^{-1}\partial^2_{t} g^{(0)}) -z^2{\rm Tr}
(h^{-1}g^{(1)}) 
\cr&& \cr&& \;\;\;\;\;\;\;\;\;\;\;\;\;\;
+\frac{1}{2}z(z-\delta){\rm Tr}((h^{-1}\partial_t g^{(0)})^2)+\frac{z^2}{2}(d-1)g_{tt}^{(1)}+
\cdots \bigg)\, ,
\eea
as well as
\bea
\Theta\cdot \Theta=\Theta^a_{b}\Theta^b_{a} &=& \frac{\alpha^2z^2}{L^4}\bigg((d-1)
-z{\rm Tr}(h^{-1}\partial_t g^{(0)})-z(z-\delta){\rm Tr}(h^{-1}\partial^2_{t} g^{(0)})
\cr&& \cr &&
\;\;\;\;\;\;\;\;\;\;\;\;\;\;\;
-2z^2{\rm Tr}(h^{-1}g^{(1)})
+z(z-\delta){\rm Tr}((h^{-1}\partial_t g^{(0)})^2)
\cr && \cr &&
\;\;\;\;\;\;\;\;\;\;\;\;\;\;\; 
+\frac{z^2}{4}{\rm Tr}((h^{-1}\partial_t g^{(0)})^2)+z^2(d-1)g_{tt}^{(1)}+\cdots \bigg).
\label{Theta.Theta}
\eea
Moreover, if one applies eq. \eqref{qs}, the above expressions may be recast into the following forms
\bea\label{TTT}
\Theta &=&\frac{\alpha z}{L^2}\bigg((d-1)-zq_{1}^{(0)}-z(z-\delta)(2q_{2}^{(0)}-
(q_{1}^{(0)})^2)-2z^2q_{0}^{(2)}+\frac{z^2}{2}(d-1)g_{tt}^{(1)}+\cdots \bigg),\nonumber\\
\Theta\cdot \Theta &=& \frac{\alpha^2z^2}{L^4}\bigg((d-1)-2zq_{1}^{(0)}+(3z^2-2z\delta)
(q_{1}^{(0)})^2-4z(z-\delta)q_{2}^{(0)}+4(d-3)z^2q_{0}^{(2)}\cr &&\cr
&&\;\;\;\;\;\;\;\;\;\;\;\;\;\;\;\;\;\;\;\;\;\;\;\;\;-z^2{\cal R}+z^2(d-1)
g_{tt}^{(1)} +\cdots
\bigg)\,\, .
\eea
We have now all the ingredients to compute the on-shell action and find the corresponding divergent terms. To proceed, we will only consider the contribution of different terms to the action near $r=\delta$ as shown in figure \ref{fig:C}. Moreover, since the two regularizations are the same, we choose the second regularization.
Actually using the expressions we have presented so far, it is straightforward to show that for one-half of the WDW patch, the divergent part of the on-shell action to order $\mathcal{O}(\delta^{d-3})$ is given by
\bea\label{AC}
I&=&\frac{1}{16\pi G_N}\int d^{d+1}x\sqrt{-g}(R-2\Lambda)
+\frac{1}{8\pi G_N}
\int_{t=0,r=\delta}d^d\sigma\sqrt{h}\log\frac{|k_P\cdot k_F|}{2}\cr &&\cr
&& +\frac{1}{8\pi G_N} \int_{F}d\lambda d^{d-1}\sigma\sqrt{h}\Theta\log\frac{L|\Theta|}{d-1}
+\frac{1}{8\pi G_N} \int_{P}d\lambda d^{d-1}\sigma\sqrt{h}\Theta\log\frac{L|\Theta|}{d-1}
\cr &&\cr
&=&\frac{L^{d-1}}{4\pi G_{N}(d-3)\delta^{d-3}}\int d^{d-1}\sigma
\sqrt{h}\bigg(\frac{d-1}{d-2}g_{tt}^{(1)}- \frac{2(d-1)}{d-2}q_{0}^{(2)}- \frac{2q_{2}^{(0)}}{d-2}
+\frac{(q_{1}^{(0)})^2}{2(d-1)}\bigg) \nonumber\\ &&\cr
&&+{\cal O}\left(\frac{1}{\delta^{d-5}}\right)+{\rm finite\, terms}\,,
\eea  
where $P$ and $F$ stand for past and future null boundaries whose normal vectors are also denoted by $k_p$ and $k_F$, respectively.  Furthermore, by applying eq. \eqref{gg} the leading divergent term of the on-shell action is given by
\be\label{I-Div}
I=-\frac{L^{d-1}}{4\pi G_{N}(d-3)\delta^{d-3}}\int d^{d-1}\sigma
\sqrt{h}\bigg(\frac{-(q_{1}^{(0)})^2}{2(d-1)}-2q_{0}^{(2)}+\frac{1}{d-2}\cal R
\bigg)\,.
\ee
It is worth mentioning that we have already fixed the undetermined length scale $\tilde{L}$ in the counterterm $I_{\rm ct}^{(0)}$ as $\tilde{L}=\frac{L}{d-1}$. One can show that this choice removes the most divergent term of holographic complexity which is at order $\frac{1}{\delta^{d-1}}$. Therefore, the most divergent term that is remained in eq. \eqref{I-Div} is at order $\frac{1}{\delta^{d-3}}$. Now the aim is to add proper counterterms to remove these UV divergences. We note, however, that using minimal subtraction, new counterterms have been studied in \cite{Kim:2017lrw} that is essentially the above terms with a minus sign. Of course, since eventually we would like to have a covariant action, it is curtail to make sure that adding any terms would not alter the variational principle.  Therefore in what follows, we would like to introduce new counterterms defined on the null boundaries of the WDW patch which remove the above divergent terms. 
\\Actually the counterterms should be written in terms of the induced metric on the null boundary and possibly its derivative. To write the corresponding counterterms, one may apply eq. \eqref{TTT} and obtain the following expressions
\bea 
\int d\lambda d^{d-1}\sigma \sqrt{\gamma} \bigg( \frac{\Theta\cdot\Theta}{\Theta} \!\!\!  && - \frac{\Theta}
{d-1}\bigg) \!\!\!\!\!\!\!\!\!\!
\cr && \cr 
\!\!\!\!\!\!\!\!\!\!\!\!\!\!\!\!\!\! &&= \! \frac{-(d-2)L^{d-1}}{(d-1)(d-3)\delta^{d-3}}\int d^{d-1}\sigma \sqrt{h}
\bigg(\frac{(q_{1}^{(0)})^2}{d-1}+4q_{0}^{(2)}-\frac{\cal R}{d-2} \bigg),
\cr && \cr
\int d\lambda d^{d-1}\sigma \sqrt{\gamma}\,\Theta\,{\cal R}[\gamma]&&=-\frac{(d-1)}
{(d-3)}\,\frac{L^{d-3}}{\delta^{d-3}}\int d^{d-1}\sigma \sqrt{h}\,{\cal R}[g].
\eea 
It is then straightforward to see that the divergent terms in eq.\eqref{I-Div} may be written as follows
\bea
I \! =\! -\frac{1}{16\pi G_{N}} \!\! \int d\lambda d^{d-1}\sigma \sqrt{\gamma} \bigg[
\frac{d-1}{d-2} \!\left(\frac{\! \Theta\cdot\Theta}{\Theta
}\!- \! \frac{\Theta}{d-1}\right) \! - \! \frac{L^2}{(d-1)(d-2)}\Theta {\cal R}[\gamma]+\cdots
\bigg], \;\;\;
\eea  
where the integration is over future and past null surfaces of one side of the WDW patch. It is then easy to write the corresponding counterterms that are essentially the above expressions with a minus sign, i.e. 
\bea
I_{ct}^{(1)} \!\!=\!\! \frac{1}{16\pi G_{N}} \!\! \int_{\mathcal{N}} \!  d\lambda d^{d-1}\sigma \sqrt{\gamma}\bigg[
\frac{d-1}{d-2}\left(\frac{\Theta\cdot\Theta}{\Theta
} \!- \! \frac{\Theta}{d-1}\right) \!\!-\! \frac{L^2}{(d-1)(d-2)}\Theta {\cal R}[\gamma] \! + \! \cdots
\bigg], \,
\label{I-ct-1}
\eea  
and we have to calculate it for each null boundary $\mathcal{N}$ of the WDW patch. It should be pointed out that the above counterterms work for $d=3$ and 4. When, $d=2$ with the chosen value for $\tilde{L}=\frac{L}{d-1}$, the holographic complexity is finite and no counterterms are needed (see also \cite{Kim:2017lrw}). Moreover, for higher dimensions, it seems that higher powers of $\mathcal{R}$ and $\mathcal{R}_{ij}$ would appear in the above expression. Furthermore, it is evident that for black branes these null counterterms are zero. In the next section, we compute the above counterterms for an AdS-Schwarzschild black hole. 

\section{Holographic complexity}
\label{sec: Holographic Complexity}

In this section, we calculate the holographic complexity for the AdS-Schwarzschild solution \eqref{metric-AdS-Schwarzschild} at time $t>t_c$. As mentioned above, the two methods of regularization are the same, hence we apply the second one. In this case, the WDW patch is given by the right panel of figure \ref{fig:C0}. The bulk action is as follows
\bea
I_{\rm bulk} = 
-\frac{d V_{d-1} L^{2(d-1)}}{4 \pi G_N}  
\bigg[ 
&& \!\!\!\! 2 \int_{\delta}^{r_{\rm max}} \frac{dr}{r^{d+1}}
 (r^*(\delta) - r^*(r)) 
\cr && \cr && \;\;\;\;\;\;\;\;\;\;\;\;\; + \int_{r_m}^{r_{\rm max}} \frac{dr}{r^{d+1}}(t_R - r^*(\delta) + r^*(r) )
 \bigg].
\eea 
Then, the above integrals can be rewritten as follows
\bea
I_{\rm bulk} = \frac{V_{d-1} L^{2(d-1)}}{4 \pi G_N}  \bigg[ && \!\!\!\! -2  \int_{\delta}^{r_{\rm m}} \frac{dr}{r^d}  \frac{1}{f(r)}
- \int_{r_m}^{r_{\rm max}} \frac{dr}{r^d} \frac{1}{f(r)} 
\cr && \cr &&
\;\;\;\;\;\;\;\;\;\;\;\; +\frac{1}{r_{\rm max}^d} \left( \frac{t}{2}+r^*(\delta) - r^*(r_{\rm max}) \right) \bigg].
\label{I-bulk-2}
\eea
Now by integration by parts, the bulk action can be recast into
\bea
&&I_{\rm bulk} = \frac{V_{d-1} L^{2(d-1)}}{4 \pi G_N}  \bigg[ \frac{1}{r_0^d}\left( (r^*(r_m)-r^*(\delta) )+ (r^*(r_{\rm max}) - r^*(\delta))\right)
\cr&& \cr && 
\;\;\;\;\;\;\;\;\;\;\;\;\;\;\;\;\;\;\;\;\;\;\;\;\;\;
+ \frac{1}{(d-1)}\left(-\frac{2}{\delta^{d-1}}+\frac{1}{r_{\rm max}^{d-1}}+\frac{1}{r_m^{d-1}} \right)
+ \frac{2}{L^2} \int_{\delta}^{r_m} \frac{dr}{r^{d-2}} +\frac{1}{L^2} \int_{r_m}^{r_{\rm max}} \frac{dr}{r^{d-2}}
\cr&& \cr&&
\;\;\;\;\;\;\;\;\;\;\;\;\;\;\;\;\;\;\;\;\;\;\;\;\;\;
 -\frac{2}{L^4} \int_{\delta}^{r_m} \frac{dr}{r^{d-4} f(r)} - \frac{1}{L^4} \int_{r_m}^{r_{\rm max}} \frac{dr}{r^{d-4} f(r)} 
+ \frac{2}{L^2 r_0^d} \int_{\delta}^{r_m} \frac{dr \; r^2}{f(r)}
\cr&& \cr&&
 \;\;\;\;\;\;\;\;\;\;\;\;\;\;\;\;\;\;\;\;\;\;\;\;\;\;
 + \frac{1}{L^2 r_0^d} \int_{r_m}^{r_{\rm max}} \frac{dr \; r^2}{f(r)} 
+\frac{1}{r_{\rm max}^d} \left( \frac{t}{2}+r^*(\delta) - r^*(r_{\rm max}) \right) \bigg].
\label{I-bulk-3}
\eea
There is a GHY term for the future singularity at $r=r_{\rm max}$,
\bea
I_{\rm GHY}&=&  \frac{V_{d-1} L^{2(d-1)}}{8 \pi G_N}\frac{1}{r^d} \left( -2 d f(r) + r f'(r)  \right) (t_R - r^*(r_{\rm max}) + r^*(\delta) ) |_{r_{\rm max}} 
\cr && \cr
&=& \frac{V_{d-1} L^{2(d-1)}}{4 \pi G_N} \left(-\frac{d}{r_{\rm max}^d}+ \frac{(1-d)}{L^2 r_{\rm max}^{d-2}} +\frac{d}{2r_0^d} \right)(\frac{t}{2}- r^*(r_{\rm max}) + r^*(\delta) ).
\label{I-GH-1}
\eea 
By adding it to the bulk action, one has
\bea
I_{\rm bulk} + I_{\rm GHY} &=& \frac{V_{d-1} L^{2(d-1)}}{4 \pi G_N}  \bigg[ \frac{1}{(d-1)}\left(-\frac{2}{\delta^{d-1}}+\frac{1}{r_{\rm max}^{d-1}}+\frac{1}{r_m^{d-1}} \right)
\cr && \cr &&
\;\;\;\;\;\;\;\;\;\;\;\;\;\;
+ \frac{2}{L^2} \int_{\delta}^{r_m} \frac{dr}{r^{d-2}} +\frac{1}{L^2} \int_{r_m}^{r_{\rm max}} \frac{dr}{r^{d-2}}
-\frac{2}{L^4} \int_{\delta}^{r_m} \frac{dr}{r^{d-4} f(r)} 
\cr&& \cr&&
\;\;\;\;\;\;\;\;\;\;\;\;\;\;
- \frac{1}{L^4} \int_{r_m}^{r_{\rm max}} \frac{dr}{r^{d-4} f(r)} 
+ \frac{2}{L^2 r_0^d} \int_{\delta}^{r_m} \frac{dr \; r^2}{f(r)} + \frac{1}{L^2 r_0^d} \int_{r_m}^{r_{\rm max}} \frac{dr \; r^2}{f(r)} 
\cr&& \cr&&
\;\;\;\;\;\;\;\;\;\;\;\;\;\;
+ \left( \frac{2-d}{2 r_0^d}+\frac{(d-1)}{r_{\rm max}^{d-2}} \left(\frac{1}{r_{\rm max}^2}+\frac{1}{L^2}\right)\right) (r^*(r_{\rm max}) - r^*(\delta)) 
\cr&& \cr&&
\;\;\;\;\;\;\;\;\;\;\;\;\;\;
+ \frac{r^*(r_m) -r^*(\delta)}{r_0^d}  +\frac{t}{2} \left(\frac{d}{2 r_0^d} +\frac{(1-d)}{r_{\rm max}^{d-2}} \left(\frac{1}{r_{\rm max}^2} + \frac{1}{L^2}\right)\right)
\bigg].
\label{I-bulk-GH}
\eea
On the other hand, there are four joint points and their contributions are given by 
\bea
I_{\rm joints} \!=\! \frac{V_{d-1} L^{2(d-1)}}{4\pi G_N} \! \left[-\frac{1}{\delta^{d-1}} \log\frac{\alpha \beta \delta^2}{L^2 f(\delta) }+ \! \frac{1}{2 r_m^{d-1}} \log\frac{\alpha \beta r_m^2}{L^2 f(r_m)} + \! \frac{1}{2r_{\rm max}^{d-1}} \log\frac{\alpha \beta r_{\rm max}^2}{L^2 f(r_{\rm max}) }\right]. \;\;\;\;\;
\label{I-joints}
\eea 
The counterterm $I_{\rm ct}^{(0)}$ for the four null boundaries is as follows
\bea
I^{(0)}_{\rm ct} &=&  \frac{V_{d-1} L^{2(d-1)}}{2 \pi G_N} \frac{1}{\delta^{d-1}} \left[ \frac{1}{(d-1)} + \log \frac{ \sqrt {\alpha \beta}  \delta}{L} \right]
\cr && \cr&& \;  
- \frac{V_{d-1} L^{2(d-1)}}{4 \pi G_N} \frac{1}{r_m^{d-1}} \left[ \frac{1}{(d-1)} + \log \frac{ \sqrt{\alpha \beta}  r_m}{L} \right]  
\cr && \cr&& \;  
- \frac{V_{d-1} L^{2(d-1)}}{4 \pi G_N} \frac{1}{r_{\rm max}^{d-1}} \left[ \frac{1}{(d-1)} + \log \frac{ \sqrt{\alpha \beta}  r_{\rm max}}{L} \right].
\label{I-ct0}
\eea 
From eq. \eqref{I-joints} and eq. \eqref{I-ct0}, it is evident that the ambiguities $\alpha$ and $\beta$ are canceled in the action, and hence one can write 
\bea
I &=& I_{\rm bulk} + I_{\rm GHY} + I_{\rm joints} + I^{(0)}_{\rm ct} 
\cr&&\cr
&=& \frac{V_{d-1} L^{2(d-1)}}{4 \pi G_N}  \bigg[ 
 \frac{2}{L^2} \int_{\delta}^{r_m} \frac{dr}{r^{d-2}} +\frac{1}{L^2} \int_{r_m}^{r_{\rm max}} \frac{dr}{r^{d-2}}
\cr&&\cr&& \;\;\;\;\;\;\;\;\;\;\;\;\;\;\;\;\;\;\;\;\;\;\;
-\frac{2}{L^4} \int_{\delta}^{r_m} \frac{dr}{r^{d-4} f(r)} - \frac{1}{L^4} \int_{r_m}^{r_{\rm max}} \frac{dr}{r^{d-4} f(r)} 
\cr&&\cr&& \;\;\;\;\;\;\;\;\;\;\;\;\;\;\;\;\;\;\;\;\;\;\;
+ \frac{2}{L^2 r_0^d} \int_{\delta}^{r_m} \frac{dr \; r^2}{f(r)} + \frac{1}{L^2 r_0^d} \int_{r_m}^{r_{\rm max}} \frac{dr \; r^2}{f(r)} 
\cr&&\cr&& \;\;\;\;\;\;\;\;\;\;\;\;\;\;\;\;\;\;\;\;\;\;\;
+ \left( \frac{2-d}{2 r_0^d}+\frac{(d-1)}{r_{\rm max}^{d-2}} \left(\frac{1}{r_{\rm max}^2}+\frac{1}{L^2}\right)\right) (r^*(r_{\rm max}) - r^*(\delta)) 
\cr&&\cr&& \;\;\;\;\;\;\;\;\;\;\;\;\;\;\;\;\;\;\;\;\;\;\;
+ \frac{r^*(r_m) -r^*(\delta)}{r_0^d}  +\frac{t}{2} \left(\frac{d}{2 r_0^d} +\frac{(1-d)}{r_{\rm max}^{d-2}} \left(\frac{1}{r_{\rm max}^2} + \frac{1}{L^2}\right)\right)
\cr && \cr && \;\;\;\;\;\;\;\;\;\;\;\;\;\;\;\;\;\;\;\;\;\;\;
+ \frac{1}{\delta^{d-1}} \log f(\delta) - \frac{1}{2 r_m^{d-1}} \log f(r_m) - \frac{1}{2 r_{\rm max}^{d-1}} \log f(r_{\rm max}) 
\bigg].
\label{I-bulk-GH-jnt+ct0-1}
\eea
Now to extract the UV divergent terms in eq. \eqref{I-bulk-GH-jnt+ct0-1}, we expand it around $\delta=0$. When d is even, we have
\bea
I &=& \frac{V_{d-1} L^{2(d-1)}}{4 \pi G_N}  \bigg[ 
\frac{(d-1)}{(d-3) } \frac{1}{L^2 \delta^{d-3}} - \frac{1}{(d-3) L^2}\left( \frac{1}{r_{\rm max}^{d-3}}+ \frac{1}{r_m^{d-3}} \right)
\cr && \cr &&\;\;\;\;\;\;\;\;\;\;\;\;\;\;\;\;\;\;\;\;\;\;\;
-\frac{2}{L^4} \int_{\delta}^{r_m} \frac{dr}{r^{d-4} f(r)} - \frac{1}{L^4} \int_{r_m}^{r_{\rm max}} \frac{dr}{r^{d-4} f(r)} 
\cr && \cr &&\;\;\;\;\;\;\;\;\;\;\;\;\;\;\;\;\;\;\;\;\;\;\;
+ \frac{2}{L^2 r_0^d} \int_{\delta}^{r_m} \frac{dr \; r^2}{f(r)} + \frac{1}{L^2 r_0^d} \int_{r_m}^{r_{\rm max}} \frac{dr \; r^2}{f(r)} 
\cr && \cr &&\;\;\;\;\;\;\;\;\;\;\;\;\;\;\;\;\;\;\;\;\;\;\;
+ \left( \frac{(2-d)}{2 r_0^d}+\frac{(d-1)}{r_{\rm max}^{d-2}} \left(\frac{1}{r_{\rm max}^2}+\frac{1}{L^2}\right)\right) (r^*(r_{\rm max}) - r^*(\delta)) 
\cr && \cr &&\;\;\;\;\;\;\;\;\;\;\;\;\;\;\;\;\;\;\;\;\;\;\;
+ \frac{r^*(r_m) -r^*(\delta)}{r_0^d}  +\frac{t}{2} \left(\frac{d}{2 r_0^d} +\frac{(1-d)}{r_{\rm max}^{d-2}} \left(\frac{1}{r_{\rm max}^2} + \frac{1}{L^2}\right)\right)
\cr && \cr &&\;\;\;\;\;\;\;\;\;\;\;\;\;\;\;\;\;\;\;\;\;\;\;
-  \frac{1}{2 r_m^{d-1}} \log f(r_m) - \frac{1}{2 r_{\rm max}^{d-1}} \log f(r_{\rm max}) + \cdots
\bigg].
\label{I-bulk-GH-jnt+ct0-2}
\eea
Now we calculate the new counterterm $I^{(1)}_{\rm ct} $ introduced in eq. \eqref{I-ct-1}. The induced metric on the null surfaces is given by
\bea
ds^2_{\mathcal{N}}=\frac{L^4}{r^2}d \Omega_{d-1},
\eea 
then for the null boundary $B^\prime_1$, from eq. \eqref{null-vectors}, \eqref{Theta-ab} and \eqref{Theta.Theta},  one obtains
\bea
\mathcal{R}&=&\frac{(d-1)(d-2)r^2}{L^4} ,
\nonumber\\
\Theta^a_b &=& \frac{\alpha \; r}{L^2} \delta^a_b,
\nonumber\\
\Theta.\Theta &=&\frac{(d-1) \alpha^2 r^2}{L^4}.
\eea 
Therefore, one has
\bea
\frac{\Theta.\Theta}{\Theta}-\frac{\Theta}{d-1}=0,
\label{Theta term-Null counterterm}
\eea 
and the first term in eq. \eqref{I-ct-1} vanishes. In other words, for each null boundary $I_{\rm ct}^{(1)}$ is given by
\bea
I^{(1)}_{\rm ct} &=& \frac{1}{16 \pi G_N} \frac{L^2}{(d-1)(d-2)} \int_{\mathcal{N}} d \lambda d^{d-1} \Omega \sqrt{\gamma} \Theta \mathcal{R}[\gamma]
\nonumber \\
&=& -\frac{(d-1) V_{d-1} L^{2(d-2)}}{16 \pi G_N} \int_{\delta} \frac{dr}{r^{d-2}}.
\label{I-ct-1-d}
\eea 
For $d \neq 3$, the contribution of the four null boundaries is as follows
\bea
I^{(1)}_{\rm ct}  =  \frac{V_{d-1} L^{2(d-2)}}{4 \pi G_N} \frac{(d-1)}{(d-3)} \left[-\frac{1}{\delta^{d-3}}+ \frac{1}{2 r_m^{d-3}} + \frac{1}{2r_{\rm max}^{d-3}} \right].
\label{I-ct-1-d-neq to 3}
\eea 
Therefore, the total action is given by 
\bea
I_{\rm tot} = I + I_{\rm ct}^{(1)}.
\label{I-tot}
\eea
In the following, we want to study the behavior of $I_{\rm tot}$ when $\delta \rightarrow 0$ and $r_{\rm max} \rightarrow \infty$. Since, our null counterterms \eqref{I-ct-1}, are valid for $d<5$, we consider the cases for which $d=3,4$.

\subsection{d=4}
\label{sec: d=4}

For $d=4$, if we take the limit $r_{\rm max} \rightarrow \infty$, then eq. \eqref{I-bulk-GH-jnt+ct0-2} is simplified as follows
\bea
&&I = \frac{V_{3} L^{6}}{4 \pi G_N}  \bigg[ 
\frac{1}{L^2} \left(\frac{3}{\delta} - \frac{1}{r_m}\right) + \frac{2}{L^2 r_0^4} \int_{\delta}^{r_m} \frac{dr \; r^2}{f(r)} + \frac{1}{L^2 r_0^4} \int_{r_m}^{r_{\rm max}} \frac{dr \; r^2}{f(r)} 
\cr && \cr &&\;\;\;\;\;\;\;\;\;\;\;\;\;\;\;\;\;\;\;\;\;
+  (r^*(r_{\rm max}) - r^*(\delta)) \left( \frac{1}{L^4} -\frac{1}{r_0^4} \right) + ( r^*(r_m) -r^*(\delta)) \left(\frac{2}{L^4}+\frac{1}{r_0^4} \right) 
\cr && \cr &&\;\;\;\;\;\;\;\;\;\;\;\;\;\;\;\;\;\;\;\;\;
+  \frac{t}{r_0^4} - \frac{1}{2 r_m^{3}} \log f(r_m) 
\bigg].
\label{I-bulk-GH-jnt+ct0-d4-2}
\eea
Moreover, from eq. \eqref{I-ct-1-d} we can write the $I_{\rm ct}^{(1)}$ for the four null surfaces as follows
\bea
I^{(1)}_{\rm ct}  =  3 \frac{V_3 L^4}{4 \pi G_N} \left[-\frac{1}{\delta}+ \frac{1}{2 r_m} \right].
\label{I-ct-1-d4}
\eea 
Now one can see that the UV divergent terms in eq. \eqref{I-ct-1-d4} and eq. \eqref{I-bulk-GH-jnt+ct0-d4-2} cancel each other, and the total action,
\bea
&&I _{\rm tot}= \frac{V_{3} L^{6}}{4 \pi G_N}  \bigg[ \frac{1}{2 L^2 r_m} + \frac{2}{L^2 r_0^4} \int_{\delta}^{r_m} \frac{dr \; r^2}{f(r)} + \frac{1}{L^2 r_0^4} \int_{r_m}^{r_{\rm max}} \frac{dr \; r^2}{f(r)} 
\cr && \cr &&\;\;\;\;\;\;\;\;\;\;\;\;\;\;\;\;\;\;\;\;\;\;\;\;
+  (r^*(r_{\rm max}) - r^*(\delta)) \left( \frac{1}{L^4} -\frac{1}{r_0^4} \right) + ( r^*(r_m) -r^*(\delta)) \left(\frac{2}{L^4}+\frac{1}{r_0^4} \right) 
\cr && \cr &&\;\;\;\;\;\;\;\;\;\;\;\;\;\;\;\;\;\;\;\;\;\;\;\;
+   \frac{t}{r_0^4} - \frac{1}{2 r_m^{3}} \log f(r_m) 
\bigg],
\label{I-tot-d-4}
\eea
is convergent in the  limit $r_{\rm max} \rightarrow \infty$. Therefore, the counterterm introduced in eq. \eqref{I-ct-1} removes all the UV divergences of the on-shell action. Next, one can substitute the tortoise coordinate $r^*(r)$ in eq. \eqref{I-tot-d-4} and take the integrals, to find the finite part of the on-shell action explicitly. To do so, it is fruitful to decompose the $\frac{1}{f(r)}$ factor in eq.  \eqref{tortoise} as follows
\bea
\frac{1}{f(r)} = \frac{L^2 r_h}{2(L^2 + 2 r_h^2)(r-r_h)} + \frac{L^2  \left( L^2 (2 r +r_h) + r_h (-r^2 + 2 r r_h + r_h^2 \right)}{2 (r+r_h)(L^2 + r^2 + r_h^2)(L^2+ 2 r_h^2)}.
\eea 
Then, it is straightforward to calculate the tortoise coordinate as follows
\bea
r^*(r) = \frac{L^2 r_h}{2( 2 L^2+ r_h^2)}  \log \frac{|r-r_h |}{r+r_h} + \frac{L r_h \sqrt{L^2 + r_h^2}}{2 L^2 + r_h^2} \tan^{-1} \left(\frac{L r_h}{z \sqrt{L^2+ r_h^2}}\right).
\label{r*d=4}
\eea 
 Plugging this expression into eq. \eqref{I-tot-d-4}, one observes that the result is too much complicated to be illuminating. However, one can consider the limit in which the horizon radius is much larger than the AdS radius. In our coordinate system where the AdS boundary is located at $r=\delta$ this happens whenever one has 
$\frac{r_h}{L} = \alpha \ll 1$.
\footnote{Note that our radial coordinate is related to the radial coordinate of \cite{Chapman:2016hwi} as follows $r_{\rm here} =\frac{L^2}{r_{\rm there}}$.}
By Taylor expansion to third order in $\alpha$, one has
\bea
I_{\rm tot} &=& \frac{V_3 L^2}{4 \pi G_N} \left( \frac{1+\alpha^2}{\alpha^4} - \frac{L^4}{2 r_m^4} \right)  t
+ \frac{3 V_3 L^4}{8 \pi G_N r_m} 
\cr&& \cr&&
 -\frac{V_3 L^3}{8 G_N \alpha} + \frac{V_3 L^6}{2 \pi G_N} \frac{1}{r_m^3} \left(- \frac{1}{6} + \frac{1}{2} \log \frac{L \alpha}{r_m}\right) 
+\cdots.
\label{I-tot-t>tc-1}
\eea 
At the end, one can trade the parameter $\alpha$ with the thermal entropy of the black hole, 
\bea
S= \frac{V_{d-1} L^{2(d-1)}}{4 G_N r_h^{d-1}} = \frac{V_{d-1}}{4 G_N} \left( \frac{L}{\alpha} \right)^{d-1},
\label{entropy}
\eea 
and write the holographic complexity as follows
\bea
\mathcal{C}= \frac{I_{\rm tot}}{\pi} &=&  \frac{2 \Upsilon}{\pi^2} \bigg[ \frac{t}{2L}\left[ \left(\frac{S}{\Upsilon}\right)^{\frac{2}{3}} + \left(\frac{S}{\Upsilon}\right)^{\frac{4}{3}} - \frac{1}{2} \left(\frac{L}{r_m}\right)^4 \right]
\cr&& \cr&& \;\;\;\;\;\;\;\;\;\;\;\;\;\;\;
+\frac{1}{8}\left(\frac{L}{r_m}\right)^4 \left(\frac{\Upsilon}{S}\right)^\frac{1}{3}+ \frac{3}{4} \frac{L}{r_m}- \frac{\pi}{4} \left(\frac{S}{\Upsilon}\right)^{\frac{1}{3}} + \frac{\pi}{8} \frac{\Upsilon}{S}
\cr&& \cr&& \;\;\;\;\;\;\;\;\;\;\;\;\;\;\;
+\left(\frac{L}{r_m}\right)^3 \left(-\frac{1}{6} + \log \frac{L}{r_m} - \frac{1}{3} \log \frac{S}{\Upsilon}\right)+ \cdots
\bigg],
\eea 
where we have defined 
\bea
\Upsilon=\frac{V_3 \pi^3}{20} C_T,
\label{Upsilon}
\eea
in terms of the central charge $C_T= \frac{5}{\pi^3} \frac{L^3}{G_N}$ of the dual CFT \cite{Buchel:2009sk}. At early times when the the joint point $m$ touches the past singularity, i.e. $r_m \rightarrow r_{ \rm max}$, one has
\bea
\mathcal{C} = \frac{2 \Upsilon}{\pi^2} \bigg[ \frac{t}{2L}\left[ \left(\frac{S}{\Upsilon}\right)^{\frac{2}{3}} + \left(\frac{S}{\Upsilon}\right)^{\frac{4}{3}} \right] - \frac{\pi}{4} \left(\frac{S}{\Upsilon}\right)^{\frac{1}{3}} + \frac{\pi}{8} \frac{\Upsilon}{S} + \cdots\bigg].
\label{complexity-BH-d=4-t>tc}
\eea 
It is evident that the above series does not look consistent with the expression for the complexity of formation in ref. \cite{Chapman:2016hwi}. We should remind the reader that the above expression is calculated for times $t>t_c$, and hence to compare it with the complexity of formation, one has to add the contribution of the GHY action (See eq. \eqref{GHY-BH-t<tc}) on the past singularity to eq. \eqref{complexity-BH-d=4-t>tc}. Moreover, one should also calculate the finite part of the holographic complexity of a global AdS spacetime (See appendix \ref{App-B}) and subtract it from eq. \eqref{complexity-BH-d=4-t>tc} according to eq. \eqref{compleity-formation}. Then it is straightforward to verify that eq. \eqref{complexity-BH-d=4-t>tc} is consistent with the complexity of formation. To clarify the issue we calculate the complexity of formation in section \ref{sec: Complexity of Formation}.

\subsection{d=3}
\label{d=3}
Now we study the case of $d=3$. When  $r_{\rm max} \rightarrow \infty$, from eq. \eqref{I-bulk-GH-jnt+ct0-1} we have
\bea
&&I = \frac{V_2 L^4}{4 \pi G_N}  \bigg[ 
\frac{1}{L^2} \left(1 -2\log \delta + \log r_m + \log r_{\rm max}\right)
\cr&& \cr&& \;\;\;\;\;\;\;\;\;\;\;\;\;\;\;\;\;\;\;\;\;
-\frac{2}{L^4} \int_{\delta}^{r_m} \frac{r\; dr}{ f(r)} - \frac{1}{L^4} \int_{r_m}^{r_{\rm max}} \frac{r \; dr}{f(r)} 
\cr&& \cr&& \;\;\;\;\;\;\;\;\;\;\;\;\;\;\;\;\;\;\;\;\;
+ \frac{2}{L^2 r_0^3} \int_{\delta}^{r_m} \frac{dr \; r^2}{f(r)} + \frac{1}{L^2 r_0^3} \int_{r_m}^{r_{\rm max}} \frac{dr \; r^2}{f(r)} 
\cr&& \cr&& \;\;\;\;\;\;\;\;\;\;\;\;\;\;\;\;\;\;\;\;\;
- \frac{(r^*(r_{\rm max}) - r^*(\delta)) }{2 r_0^3} + \frac{(r^*(r_m) -r^*(\delta))}{r_0^3} 
\cr&& \cr&& \;\;\;\;\;\;\;\;\;\;\;\;\;\;\;\;\;\;\;\;\;
+  \frac{3 t}{4  r_0^3} - \frac{1}{2 r_m^2} \log f(r_m) 
\bigg].
\label{I-bulk-GH-jnt+ct0-d3-2}
\eea
The only UV divergent term in the above expression is $\log \delta$. On the other hand, from eq. \eqref{I-ct-1-d} the new counterterm for $d =3$ is given by 
\bea
I^{(1)}_{\rm ct}  = - \frac{V_2 L^2}{4 \pi G_N} \left[ \log r_{\rm max} + \log r_m - 2 \log \delta \right].
\label{I-ct-1-3}
\eea 
From eq. \eqref{I-ct-1-3} one can easily see that all of the UV divergent terms are canceled in the total action
\bea
&& I_{\rm tot} = \frac{V_2 L^4}{4 \pi G_N}  \bigg[ 
-\frac{2}{L^4} \int_{\delta}^{r_m} \frac{r\; dr}{ f(r)} - \frac{1}{L^4} \int_{r_m}^{r_{\rm max}} \frac{r \; dr}{f(r)} 
\cr&& \cr&& \;\;\;\;\;\;\;\;\;\;\;\;\;\;\;\;\;\;\;\;
+ \frac{2}{L^2 r_0^3} \int_{\delta}^{r_m} \frac{dr \; r^2}{f(r)} + \frac{1}{L^2 r_0^3} \int_{r_m}^{r_{\rm max}} \frac{dr \; r^2}{f(r)} 
\cr&& \cr&& \;\;\;\;\;\;\;\;\;\;\;\;\;\;\;\;\;\;\;\;
- \frac{(r^*(r_{\rm max}) - r^*(\delta)) }{2 r_0^3} + \frac{(r^*(r_m) -r^*(\delta))}{r_0^3} 
\cr&& \cr&& \;\;\;\;\;\;\;\;\;\;\;\;\;\;\;\;\;\;\;\;
+  \frac{1}{L^2} +\frac{3 t}{4  r_0^3} - \frac{1}{2 r_m^2} \log f(r_m) 
\bigg].
\label{I-tot-d-3}
\eea
Therefore, the new counterterm eq. \eqref{I-ct-1} removes all the UV divergences in the on-shell action. Now one can use the following decomposition 
\bea
\frac{1}{f(r)} = \frac{L^2 r_h}{(3L^2 + r_h^2)} \left(- \frac{1}{(r-r_h)} + \frac{r_h^2 (r+r_h) + L^2 (r+2 r_h)}{r^2r_h^2 + L^2 (r^2 + r \, r_h + r_h^2)}\right),
\eea 
and calculate the tortoise coordinate as follows
\bea
r^*(r) &=& \frac{L^2 r_h}{3L^2+r_h^2} \bigg[ \log \frac{L |r-r_h |}{\sqrt{L^2 (r_h^2 + r r_h + r^2) +r^2 r_h^2}}
\cr&& \cr&&
\;\;\;\;\;\;\;\;\;\;\;\;\;\;\;\;\;  + \frac{3L^2 + 2 r_h^2}{L \sqrt{3 L^2+ 4 r_h^2}} \tan^{-1} \left(\frac{2 r r_h^2 + L^2 (2 r +r_h)}{L r_h \sqrt{3 L^2+ 4 r_h^2}}\right) \bigg].
\label{r*d=3}
\eea 
Now one can substitute eq. \eqref{r*d=3} into eq. \eqref{I-tot-d-3} and obtain the on-shell action. Similar to $d=4$, the corresponding expression is very cluttered. Therefore, we consider the case of large black holes for which, $\frac{r_h}{L} = \alpha \ll 1$. To third order in $\alpha$, one obtains
\bea
I_{\rm tot}&=& \frac{V_2 L}{8 \pi G_N} t_R \left( \frac{3(1+\alpha^2)}{\alpha^3}-\frac{2 L^3}{r_m^3} \right) + \frac{V_2 L^2}{216 \pi G_N}(45-2\sqrt{3}) 
\cr&& \cr&&
- \frac{V_2 L^2}{24\sqrt{3}G_N\alpha^2}
-  \frac{V_2 L^2}{4 \pi G_N}\log\frac{r_m r_{\rm max}}{\alpha^2 L^2}
+ \frac{V_2 L^4}{8\pi G_N}\frac{1}{r_m^2}\left(-1 + 3\log\frac{\alpha L}{r_m}\right) 
\cr&& \cr&&
+\frac{V_2 L^5}{12 \sqrt{3} G_N } \frac{\alpha}{r_m^3}
+ \frac{V_2 L^2}{64 \pi G_N}(-9+\sqrt{3 \pi}) \alpha^2
+ \cdots,
\eea 
where $t_R=t_L= \frac{t}{2}$. The holographic complexity is given by
\bea
&& \mathcal{C} = \frac{\kappa}{2 \pi^2} \bigg[\frac{3 t}{2 L} \left[ \left(\frac{S}{\kappa}\right)^{\frac{1}{2}}
+ \left(\frac{S}{\kappa}\right)^{\frac{3}{2}}- 2 \left(\frac{L}{r_m}\right)^3  \right] 
+ \frac{1}{27} \left(45- 2 \sqrt{3}\right)
\cr&& \cr&& \;\;\;\;\;\;\;\;\;\;\;\;\;\;\;\;
-\frac{\pi}{3 \sqrt{3}} \frac{S}{\kappa} - \frac{2 \pi^2}{\kappa^2} \log \frac{r_m r_{\rm max}}{L^2} - \frac{2}{\kappa} \log \frac{S}{\kappa} 
\cr&& \cr&& \;\;\;\;\;\;\;\;\;\;\;\;\;\;\;\;
+ \left(\frac{L}{r_m}\right)^2 \left(-1 +3 \log \frac{L}{r_m} - \frac{3}{2} \log \frac{S}{\kappa}\right) 
\cr&& \cr&& \;\;\;\;\;\;\;\;\;\;\;\;\;\;\;\;
+\frac{2 \pi}{3 \sqrt{3}} \left( \frac{L}{r_m}\right)^3 \left(\frac{\kappa}{S}\right)^\frac{1}{2} 
+\frac{1}{8} (-9 + \sqrt{3} \pi ) \frac{\kappa}{S}
+ \cdots
\bigg]
\eea 
where 
\bea
\kappa= \frac{V_2 \pi^3}{12} C_T,
\label{kappa}
\eea
is defined in terms of the central charge $C_T= \frac{3 L^2}{\pi^3 G_N}$ \cite{Buchel:2009sk} of the dual CFT. Moreover, at the early times, i.e. $r_m \rightarrow r_{\rm max}$, one has
\bea
&&\mathcal{C} = \frac{\kappa}{2 \pi^2} \bigg[\frac{3 t}{2 L} \left[ \left(\frac{S}{\kappa}\right)^{\frac{1}{2}}
+ \left(\frac{S}{\kappa}\right)^{\frac{3}{2}} \right] 
+ \frac{1}{27} \left(45- 2 \sqrt{3}\right)
\cr&& \cr&& \;\;\;\;\;\;\;\;\;\;\;\;\;\;
-\frac{\pi}{3 \sqrt{3}} \frac{S}{\kappa} - \frac{4 \pi^2}{\kappa^2} \log \frac{r_{\rm max}}{L} - \frac{2}{\kappa} \log \frac{S}{\kappa} 
+\frac{1}{8} (-9 + \sqrt{3} \pi ) \frac{\kappa}{S}
+ \cdots
\bigg]
\label{complexity-BH-d=3-t>tc}
\eea 
\section{Complexity of formation}
\label{sec: Complexity of Formation}
\begin{figure}
	\begin{center}
		\includegraphics[scale=1]{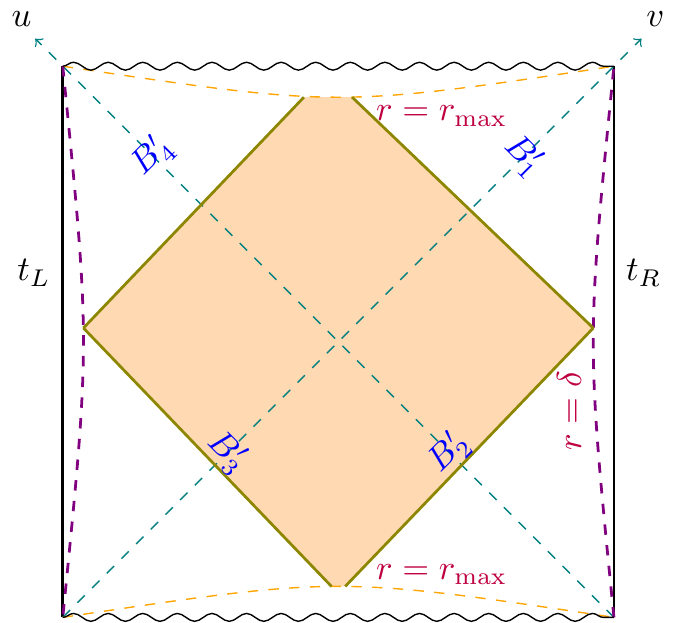}
	\end{center}
	\caption{The WDW patch for an eternal black hole at times $t < t_c$. Note that we choose the second regularization. Moreover, in contrast to figure \ref{fig:C0}, in this case, the WDW patch has an extra spacelike boundary which is a part of the past singularity.}
	\label{fig:WDW-formation}
\end{figure}

As we have seen in the last section, the null counterterms eq. \eqref{I-ct-1} are able to remove all the UV divergent terms of holographic complexity. Now one can ask whether there is any relation between this finite holographic complexity and the complexity of formation, i.e. eq. \eqref{compleity-formation}.
\footnote{We would like to thank the referee for raising the question.}
To address the question, in this section we calculate the complexity of formation for an AdS-Schwarzschild black hole for $d=2,3,4$ and compare our results with those of  \cite{Chapman:2016hwi}. To do so, first we should calculate the holographic complexity of the black hole for times $t<t_c$. Second, we should compute the holographic complexity of global AdS. In this section, we just report the holographic complexity of the black hole and postpone that of global AdS to appendix \ref{App-B}.
\\For times $t<t_c$ the Penrose diagram of a two-sided AdS-Schwarzschild black hole is shown in figure \ref{fig:WDW-formation}. It is straightforward to show that the bulk action is given by
\bea
I_{\rm bulk } = -\frac{d V_{d-1} L^{2(d-1)}}{2 \pi G_N} \int_{\delta}^{r_{\rm max}} \frac{d r}{r^{d+1}} \left( r^* (\delta) - r^*(r) \right).
\label{I-bulk-form}
\eea 
There is a GHY term for the future and past singularity, which is as follows
\bea
I_{\rm GHY} = 2 \times \frac{V_{d-1} L^{2(d-1)}}{8 \pi G_N}\frac{1}{r^d} \left(-2 d f(r) + r f'(r)\right) \left(\frac{t}{2}- r^*(r_{\rm max}) + r^*(\delta) ) \right)|_{r_{\rm max}}.
\label{GHY-BH-t<tc}
\eea 
The contributions of the four null-spacelike joints to the action are zero. Moreover, there are two null-null joints at $r=r_{\rm max}$ whose contributions to the action are as follows
\bea
I_{\rm joint} =  - \frac{V_{d-1} L^{2(d-1)}}{4 \pi G_N} \frac{1}{\delta^{d-1}} \left[ \log \frac{\alpha \beta \delta^2}{L^2} - \log f(\delta) \right].
\label{I-joint-form}
\eea
On the other hand, for the counterterm $I_{\rm ct}^{(0)}$ one obtains
\bea
I^{(0)}_{\rm ct} = \frac{V_{d-1} L^{2(d-1)} }{4 \pi G_N} \frac{1}{\delta^{d-1}} \left[ \frac{2}{(d-1)} + \log \frac{\alpha \beta \delta^2}{L^2}\right].
\label{I-ct-0-form}
\eea
Moreover, in this case all of the null boundaries of the WDW patch are extended from $r=\delta$ to $r=r_{\rm max}$. Therefore, from eq. \eqref{I-ct-1-d} one can find the null counterterm $I_{\rm ct}^{(1)}$ as follows
\begin{itemize}
	\item  for $d \neq 3$, one has
	\bea
	I^{(1)}_{\rm ct} = - \frac{V_{d-1} L^{2(d-2)} }{4 \pi G_N} \frac{(d-1)}{(d-3)} \frac{1}{\delta^{d-3}},
	\label{I-ct-1-form-d nq 3}
	\eea 
	\item for $d=3$, one obtains
	\bea
	I^{(1)}_{\rm ct} = - \frac{V_2 L^2}{2 \pi G_N} \log \frac{r_{\rm max}}{\delta}.
	\label{I-ct-1-form-d=3}
	\eea 
\end{itemize}
Now one can substitute the tortoise coordinate and calculate the complexity of the black hole as
\bea
\mathcal{C}_{\rm BH}= \frac{1}{\pi}  I_{\rm tot} =\frac{1}{\pi} \left(I_{\rm bulk} + I_{\rm GHY} + I_{\rm joints} + I_{\rm ct}^{(0)} + I_{\rm ct}^{(1)}\right)
\eea 
Similar to the $t>t_c$ case, the final expression is very complicated. Therefore, we consider large black holes, i.e. $\alpha = \frac{r_h}{L} \ll 1$ and expand the holographic complicity in powers of thermal entropy $S$ of the black hole as well as the central charge $C_T$ of the dual CFT. 

\subsection{d=2, BTZ black hole}

For $d=2$, one has $f(r) = 1- \frac{r^2}{r_h^2}$, and the solution is a non-rotating BTZ black hole. The tortoise coordinate is easily obtained 
\bea
r^*(r) = \frac{r_h}{2} \log \frac{\lvert  r- r_h \rvert }{r + r_h}.
\eea 
Moreover, the null counterterms $I_{\rm ct}^{(1)}$ are zero. It is straightforward to show that
\bea
\mathcal{C}_{\rm BH} = 0.
\label{complexity-BH-d=2-t<tc}
\eea 
In other words, the holographic complexity of a BTZ black hole has no finite term. On the other hand, the holographic complexity of a global $AdS_3$ is calculated in eq. \eqref{complexity-AdS-d=2} in appendix \ref{App-B},
\bea
\mathcal{C}_{{\rm AdS}_{3}}= \frac{V_1 L}{8 \pi G_N} =  \frac{ L}{4 G_N}.
\nonumber
\eea 
Putting everything together, one can find the complexity of formation of a BTZ black hole as follows
\bea
\Delta \mathcal{C} = - \frac{L}{2 G_N} = - \frac{C_T}{3}.
\label{DeltaC-d=2}
\eea 
In the last equality, we have used the central charge of the dual CFT, i.e. $C_T= \frac{3 L}{2 G_N}$ \cite{Brown:1986nw}. This is exactly the same value that has been obtained in eq. (4.8) in ref. \cite{Chapman:2016hwi}.  Therefore, the addition of the null counterterms eq. \eqref{I-ct-1} do not change the complexity of formation. As we will see in the following, this will happen in higher dimensions.

\subsection{d=3}

For $d=3$, the tortoise coordinate is given by eq. \eqref{r*d=3} and one has
\bea
I_{\rm tot} &=&  \frac{V_2 L^2}{12 \sqrt{3} G_N \alpha^2} + \frac{V_2 L^2}{54 \pi G_N} \left(18+ \sqrt{3} \pi +27 \log \frac{L}{r_{\rm max}} +27 \log \alpha\right) 
\cr&& \cr&&
- \frac{V_2 L^2}{54 \pi G_N} (9 - \sqrt{3} \pi ) \alpha^2+ \mathcal{O}(\alpha^3).
\eea 
Now by applying eq. \eqref{entropy} and eq. \eqref{kappa}, one can rewrite it as follows
\bea
\mathcal{C}_{\rm BH} &=& \frac{1}{3 \sqrt{3} \pi} S + \frac{4 \kappa}{54 \pi^2} \left(18+ \sqrt{3} \pi + 27 \log \frac{L}{r_{\rm max}} - \frac{27}{2} \log \frac{S}{\kappa}\right)
\cr&& \cr&&
-\frac{4 \kappa^2}{54 \pi^2} \left(9 - \sqrt{3} \pi \right) \frac{1}{S} + \cdots.
\label{complexity-BH-d=3-t<tc}
\eea 
On the other hand, the holographic complexity of a global $AdS_4$ is calculated in appendix \ref{App-B} (See eq. \eqref{complexity-AdS-d=3})
\bea
\mathcal{C}_{{\rm AdS}_{4}}= \frac{V_2 L^2}{8 \pi^2 G_N} \left( 1 + 2 \log \frac{L}{r_{\rm max}}\right).
\nonumber
\eea 
Now one can find the complexity of formation, i.e. \eqref{compleity-formation},  of a four dimensional AdS-Schwarzschild as follows
\bea
\Delta \mathcal{C} &=&  \frac{S}{3 \sqrt{3} \pi} + \frac{\pi V_2}{324} C_T \left(9+ 2 \sqrt{3} \pi \right)
- \frac{V_2 \pi}{12} C_T \log \left( \frac{12}{V_2 \pi^3} \frac{S}{C_T}\right) 
\cr&& \cr&&
-\frac{V_2^2 \pi^4}{1944} \frac{C_T^2}{S} (9 - \sqrt{3} \pi ) + \cdots,
\label{DeltaC-d=3}
\eea 
which is exactly the same as eq. (3.26) in ref. \cite{Chapman:2016hwi}. Moreover, the IR divergent term, i.e. 
$\log r_{\rm max}$, in eq. \eqref{complexity-BH-d=3-t<tc} is canceled by the same term in eq. \eqref{complexity-AdS-d=3}.

\subsection{d=4}

For $d=4$, the tortoise coordinate is given by eq. \eqref{r*d=4} and one obtains 
\bea
I_{\rm tot} =  \frac{V_3 L^3}{4 G_N} \frac{ (L^4 - r_h^4) \sqrt{L^2+r_h^2} }{r_h^3 (2 L^2 + r_h^2)}.
\label{I-d=4,t<tc}
\eea 
When $\frac{r_h}{L}=\alpha \ll 1$, one has
\bea
I_{\rm tot} =  \frac{V_3 L^3}{8 G_N \alpha^3} - \frac{9 V_3 L^3 \alpha }{64 G_N}+ \frac{V_3 L^3 \alpha^3}{64 G_N} + \cdots,
\eea 
which one can rewrite it in terms of the entropy of the black hole eq. \eqref{entropy} and the constant $\Upsilon$ in eq. \eqref{Upsilon} as follows
\bea
\mathcal{C}_{\rm BH} =  \frac{S}{2 \pi} - \frac{9 \Upsilon}{16 \pi} \left(\frac{\Upsilon}{S} \right)^{\frac{1}{3}} +\frac{\Upsilon^2}{16 \pi} S + \cdots.
\label{complexity-BH-d=4-t<tc}
\eea 
On the other hand, the holographic complexity of a global $AdS_5$ is given by eq. \eqref{complexity-AdS-d=4},
\bea
\mathcal{C}_{{\rm AdS}_{5}}=- \frac{V_3 L^3}{8 \pi G_N}.
\nonumber
\eea 
Therefore the complexity of formation is given by
\bea
\Delta \mathcal{C} &=& \frac{V_3 L^3}{4 \pi G_N} \left[ 1+ \frac{(L^2 - r_h^2) (L^2+r_h^2)^{\frac{3}{2}}}{r_h^3 (2L^2 + r_h^2)}\right].
\label{DeltaC-d=4}
\eea 
This formula agrees with eq. (3.14) in \cite{Chapman:2016hwi}. In the large horizon limit, one has
\bea
\Delta \mathcal{C}= \frac{V_3 L^3}{4 \pi G_N} + \frac{S}{2 \pi} - \frac{9}{16 \pi } \left(\frac{V_3 L^3}{4 G_N} \right)^{\frac{1}{3}} \frac{1}{S^\frac{4}{3}}+ \frac{1}{\pi }\left( \frac{V_3 L^3 }{16 G_N} \right)^2 \frac{1}{S} + \cdots,
\eea 
which is also the same as eq. (3.16) in \cite{Chapman:2016hwi}.
\section{New counterterm on the singularity}
\label{sec: New Counterterm on the Singularity}
Now we look at the behavior of the total on-shell action at times in the limit of $r_{\rm max} \rightarrow \infty$. Since, in the third line of eq. \eqref{I-bulk-GH-jnt+ct0-d3-2} we have
\bea
\int^{r_{\rm max}} \frac{r^2 dr}{f(r)} = -r_0^3 \log r_{\rm max} \; ,
\label{log-divergence}
\eea 
this log-term cancels the $log \; r_{\rm max}$ term in the first line of eq. \eqref{I-bulk-GH-jnt+ct0-d3-2}, and hence the action $I_{\rm tot}$ is convergent in this limit. However, in eq. \eqref{I-ct-1-3} there is such a term, and one can see that this log-term remains in the total action. Therefore, the total action is divergent when $r_{\rm max} \rightarrow \infty$, and this is very problematic. Actually, for odd $d$ there is always such a logarithmic IR divergent term (See also eq. \eqref{complexity-BH-d=3-t<tc}). It seems that one might resolve the issue by applying the proposals of \cite{Akhavan:2018wla,Alishahiha:2018swh,Hashemi:2019xeq}, in which it has been shown that the cutoff $r_{\rm max}$  near the singularity is related to the UV cutoff $\delta$ near the asymptotic boundary of spacetime. In particular, for AdS-Schwarzschild black holes one has \cite{Akhavan:2018wla}
\bea
r_{\rm max} \; \delta^2 = 2^{-\frac{4}{d}} r_h^3,
\label{cutoff-rmax-delta}
\eea 
here $r_h$ is the radius of the horizon. Moreover, It was suggested in ref. \cite{Akhavan:2018wla} that the $r= r_{\rm max}$ cutoff can be interpreted as a UV cutoff, and action counterterms were added on this surface. Motivated by this proposal, one might apply eq. \eqref{cutoff-rmax-delta}, and rewrite the logarithmic divergent term in eq. \eqref{log-divergence}, as follows
\bea
 \frac{V_2 L^2}{2 \pi G_N} \log \delta .
\eea 
Now our aim is to find a new type of counterterm which can cancel the above term. One possibility would be to add the following counterterm on each of the null-spacelike joint points,
\footnote{We would like to thank M. Alishahiha for bringing this point to our attention.}
which are the intersection of the null boundaries $B_1^\prime$ and $B_4^\prime$ with the future singularity at $r= r_{\rm max}$. These points are denoted by $e$ and $f$ in the right panel of figure \ref{fig:C0}
\bea
I_{\rm ct}^{(2)}= -\frac{L^2}{8 \pi G_N} \int_{\mathcal{J}} d^{d-1}  \Omega  \sqrt{h} R \; \log \delta\; , \;\;\;\;\;\;\;\;\;\;\;\; \text{for $d=3$}
\label{I-ct-2}
\eea 
here $\mathcal{J}$ is the null-spacelike joint point on the future singularity, $h$ is the determinant of  the induced metric on it, and $R$ is the Ricci scalar of the joint point.
\footnote{Note that in higher dimensions, one should use higher powers of Ricci scalar and Ricci tensor. For example in $d=5$, one might apply  
\bea
I_{\rm ct}^{(2)} \propto \int_{\mathcal{J}} d^{d-1}  \Omega  \sqrt{h} \;  \left(R_{ij}R^{ij} - \frac{d}{4(d-1)} R^2 \right)  \; \log \delta,
\eea
instead of eq. \eqref{I-ct-2}.}
Moreover, It should be emphasized that the new counterterm eq. \eqref{I-ct-2} breaks the diffeomorphism invariance of the action for odd d, and might introduce a type of anomaly in the dual CFT. At the moment, we have no idea about this anomaly, and one either has to find the source of the anomaly or give another recepie to resolve the issue. 
\\Another important point is that one could add some counterterms on the future singularity \cite{Akhavan:2018wla,Alishahiha:2019cib} similar to eq. \eqref{I-ct-HR}. However, in the limit of $r_{\rm max} \rightarrow \infty$ they go to zero and hence do not have any significance. Therefore, to extract the finite part of the on-shell action, one might add a counterterm on each boundary of the WDW patch: such that one applies eq. \eqref{I-ct-1} on null surfaces and eq. \eqref{I-ct-HR-1} on timelike surfaces. In this manner, one also has to add the counterterm  eq. \eqref{I-ct-2} on null-spacelike joint points for odd d.

\section{Growth rate of holographic complexity}
\label{sec: Growth Rate of Complexity}

Now we can talk about the rate of growth of holographic complexity. From eq. \eqref{null-bdy-reg-2}, one can find the location of the null-null joint point $m$, which is the intersection of the null boundaries $B^\prime_2$ and $B^\prime_3$, as follows
\bea
\frac{t}{2}=r^*(\delta)- r^*(r_m)
\eea 
here $t=t_L+t_R$, then one has
\bea
\frac{d r_m}{d t}= \frac{1}{2} f(r_m).
\label{drm-dt}
\eea 
The rate of growth of the on-shell action $I$ is obtained as follows \cite{Carmi:2017jqz},
\bea
\frac{d I}{d t} = 2 M+ \frac{(d-1)V_{d-1} L^{2(d-1)}}{16 \pi G_N} \frac{f(r_m)}{r_m^d} \log \lvert f(r_m) \rvert .
\label{dI-dt}
\eea 
Now we consider the rate of growth of null counterterms. For $d=3$, from eq. \eqref{I-ct-1-3}, one has
\bea
\frac{d I_{\rm ct}^{(1)}}{d t} = -\frac{V_2 L^2}{8 \pi G_N} \frac{f(r_m)}{r_m}.
\label{d-I-ct-1-d3}
\eea 
On the other hand, for $d=4$ from eq. \eqref{I-ct-1-d4}, one has
\bea
\frac{d I_{\rm ct}^{(1)}}{d t} = -\frac{3 V_3 L^4}{16 \pi G_N} \frac{f(r_m)}{r_m^2}.
\label{d-I-ct-1-d4}
\eea 
Putting everything together, one obtains the rate of growth of holographic complexity as follows
\bea
\frac{d \mathcal{C}}{d t} &=&  \frac{1}{\pi} \frac{d I_{\rm tot}}{d t}
\nonumber\\
&=& \frac{1}{\pi } \left[ 2 M+ \frac{V_{d-1} L^{2(d-1)}(d-1)}{16 \pi G_N} \frac{f(r_m)}{r_m^d} \left( \log \lvert f(r_m) \rvert - \frac{r_m^2}{L^2} \right) \right],
\label{d-I-tot-dt-d}
\eea 
here the second term in the parenthesis is the contribution of the null counterterms $I_{\rm ct}^{(1)}$. From the above expression, it is evident that at late times when $r_m \rightarrow r_h$, one has
\bea
\frac{d I_{\rm ct}^{(1)}}{d t} =0 \;\;\;\;\;\;\;\;\;\;\;\;\;\;\;\;\;\;  \text{for $r_m \rightarrow r_h$},
\eea
and hence the counterterm $I_{\rm ct}^{(1)}$ does not change the late time rate of growth of the holographic complexity. Another important point is that at early times when $t \rightarrow t_c$ , the rate of growth of the counterterm $I_{\rm ct}^{(1)}$ is positive, and hence it cannot correct the early time violation of the Lloyd bound observed in ref. \cite{Carmi:2017jqz}.

\section{Discussion}
\label{sec: Discussion}

In this paper, we first examined the equivalence of the two methods of regularization proposed in ref. \cite{Carmi:2016wjl} for an AdS-Schwarzschild black hole solution in Einstein gravity  (See figure \ref{fig:C0}). The two methods have been already studied in refs. \cite{Carmi:2016wjl,Reynolds:2016rvl}, and it has been proved that the structure of the UV divergences of holographic complexity are the same in both regularizations. However, their coefficients do not match on both sides. From figure \ref{fig:C0}, it is evident that in the first regularization the WDW patch has two extra timelike boundaries at $r=\delta$, in contrast to the WDW in the second regularization. Here, we observed that after adding timelike counterterms \eqref{I-ct-HR-1} which are inspired by holographic renormalization as well as the Gibbons-Hawking-York term for these extra timelike boundaries of the WDW patch, the coefficients of the UV divergences of holographic complexity in the two regularizations become exactly the same. Therefore, one can conclude that the two methods of regularization are completely equivalent.
\\Then we introduced new types of counterterms on null boundaries of the WDW patch for an asymptotically AdS spacetime in four and five dimensions which are covariant and are able to remove all the UV divergences of holographic complexity. To do so, we first studied the UV divergences of the gravitational on-shell action for an asymptotically AdS spacetime in the second regularization. This step has already been taken in ref. \cite{Carmi:2016wjl} and ref. \cite{Reynolds:2016rvl}. However, in refs. \cite{Carmi:2016wjl,Reynolds:2016rvl} the UV divergences have been written in terms of the extrinsic and intrinsic curvatures of the joint points. Since we were interested to find covariant counterterms on the null boundaries of the WDW patch, here we first rewrote the divergent terms in terms of the extrinsic curvature of the null boundaries, i.e. $\Theta$ and $\Theta.\Theta$. Next, by applying the minimal subtraction scheme, we found counterterms on the null boundaries. The result is presented in eq. \eqref{I-ct-1}. It should be pointed out that some type of counterterms were also introduced in ref. \cite{Kim:2017lrw} which remove all the UV divergences of holographic complexity. However, those counterterms are written on the joint points which are not the boundaries of the WDW patch. On the other hand, the counterterms found here are defined on the null boundaries of the WDW patch and are covariant.
 \\Furthermore, we calculated the complexity of formation for $d=2,3,4$, and observed that the null counterterms do not change the complexity of formation. We also applied the null counterterms to a global AdS in appendix \ref{App-B}, and observed that they also remove all the UV divergences of the holographic complexity for a global AdS. 
\\However, the null counterterms suffer from a $\log r_{\rm max}$ IR divergence for odd d, in which $r=r_{\rm max}$ is the location of the future singularity. This issue holds for both black holes and global AdS spacetimes. In the former, to resolve the problem we applied the proposals of \cite{Akhavan:2018wla,Alishahiha:2018swh,Alishahiha:2019cib,Hashemi:2019xeq}, in which it was argued that putting a UV cutoff at $r=\delta$, might introduce a cutoff behind the horizon and near the singularity at $r=r_{\rm max}$, such that they are related to each other by eq. \eqref{cutoff-rmax-delta}. If this proposal works, one can rewrite the $\log r_{\rm max}$ term in holographic complexity as $\log \delta$. Next, to remove this logarithmic divergent term, one might add a counterterm such as eq. \eqref{I-ct-2} on the null-spacelike joints located at $r=r_{\rm max}$ surface.
\\ Another important point is that these null counterterms modify the early time behavior of the growth rate of holographic complexity, although they become zero at late times when $r_m \rightarrow r_h$.
\\Moreover, one can consider charged black hole solutions \cite{Brown:2015lvg,Carmi:2017jqz,Cai:2016xho,Cai:2017sjv,Alishahiha:2019cib} such as a dyonic black hole in four dimensions whose holographic complexity is calculated in \cite{Goto:2018iay}. In this case, the metric is given by 
\bea
ds^2 &=&\frac{L^2}{r^2} \left(-f(r) dt^2+ \frac{1}{f(r)} dr^2 +L^2 d\Omega^2_{d-1}\right), \\
\label{metric-dyonic}
f(r) &=&1+ \frac{r^2}{L^2} - \frac{r^{d}}{r_0^{d}}+(Q_e^2 +Q_m^2) r^{2(d-1)},
\label{f(r)}
\eea
where $Q_{e,m}$ are related to the electric and magnetic charges of the black hole. For $d=3$, one can show that the divergent part of the action eq. \eqref{I-0} is logarithmic
\bea
(I_{\rm bulk} + I_{\rm joints} + I^{(0)}_{\rm ct})|_{\rm div.} = - \frac{\Omega_2 L^2}{2 \pi G_N} \log \delta.
\label{I-Dionic-bulk-jnt+ct0-div}
\eea
On the other hand, the contribution of the four null counterterms eq. \eqref{I-ct-1} is given by
\bea
I^{(1)}_{\rm ct}|_{\rm div.}  = + \frac{V_2 L^2}{4 \pi G_N}  \left[\log r_p + \log r_q - 2\log \delta \right],
\label{I-ct-1-Dionic}
\eea 
where the location of the bottom and tip of the null-null intersections of the corresponding WDW patch are shown by $r_p$ and $r_q$, respectively. Now, it is evident that the divergent part of the null counterterms in eq. \eqref{I-ct-1-Dionic} cancel the same logarithmic UV divergent term in eq. \eqref{I-Dionic-bulk-jnt+ct0-div}. Therefore, the null counterterms eq. \eqref{I-ct-1} are also able to remove all the UV divergences of the holographic complexity for charged black holes. Moreover, one can see that the causal structure of a charged black hole is such that the null counterterms do not end on the future singularity at $r=r_{\rm max}$. In other words, one always has $ r_{-} \leq r_{p,q} \leq r_{+}$, where $r_{\pm}$ are the radius of inner and outer horizons. Furthermore, at very late times, one has $r_p \rightarrow r_{+}$ and $r_q \rightarrow r_{-}$. Therefore, for charged black holes in contrast to AdS-Schwarzschild black holes there are not any logarithmic IR divergent terms, i.e. $\log r_{\rm max}$, for odd d. It is also interesting to consider the limit in which the charge is very small.
\footnote{We thank the referee for raising the interesting question.}
In this limit, one has $r_{-} \rightarrow r_{\rm max}$ and $r_{+} \rightarrow r_h$, and hence, at the late times $r_q \rightarrow r_{-} \rightarrow r_{\rm max}$ and $r_p \rightarrow r_{+} \rightarrow r_h$.
\footnote{See also the footnote on page 41 in ref. \cite{Carmi:2017jqz} for a discussion on the zero charge limit.}
Then form eq. \eqref{I-ct-1-Dionic} it is evident that we have an IR divergence for $d=3$ case, and we need to introduce a new counterterm, i.e. eq. \eqref{I-ct-2}, similar to the uncharged case. Another important point is that the null counterterms eq. \eqref{I-ct-1} change the growth rate of holographic complexity as follows
\bea
\frac{d \mathcal{C}}{d t} &=& \frac{L^4}{\pi G_N}\left[ Q_e^2 \; r +\frac{f(r)}{2r^3} \left( \frac{r^2}{L^2} - \log f(r) \right) \right]_{r_p}^{r_q}.
\eea 
We should emphasis that the first term in the parenthesis is the contribution of the null counterterms. Moreover, at late times, one has
\bea
\frac{d \mathcal{C}}{d t}|_{t \rightarrow \infty} &=& \frac{L^4 Q_e^2}{\pi G_N} \left( r_ {-} -  r_+ \right),
\label{growth rate-charged-late time}
\eea 
which is found in \cite{Carmi:2017jqz,Goto:2018iay} without the application of the null counterterms. Therefore, similar to the case of AdS-Schwarzschild black holes, the null counterterms do not change the growth rate of holographic complexity at late times. It should be pointed out that taking the zero charge limit, i.e. $Q_e \rightarrow 0$, in eq. \eqref{growth rate-charged-late time} is subtle, and one cannot naively set $Q_e=0$, and it needs to be investigated further.
\footnote{We would like to thank the referee for his/her useful comment on this point.}

\section*{Acknowledgment}
We would like to thank Mohsen Alishahiha very much for his collaboration and very useful comments. We are also very grateful to M. R. Tanhayi for his helpful comments on the manuscript.

\appendix
\section{Affine parameter}
\label{App-A}
In this appendix, we will find affine parameters for the null boundaries of the WDW patch in the Fefferman-Graham coordinates. In the following, we consider the future null boundary whose normal vector is given by
\bea
k_F = \alpha d(t-t_+),
\label{kF}
\eea 
where $t_+$ is given by eq. \eqref{null-surfaces-FG}. Now we define the affine parameter for the future null surface as follows
\be
k^{\mu}=\frac{\partial x^{\mu}(\lambda,\rho)}{\partial \lambda}.
\label{affine-coord-1}
\ee
In which $k^{\mu}=G^{\mu\nu}k_{\nu}$, and also the $\rho$ are the coordinates in the timelike or spacelike lines in the null surface. From eq. \eqref{kF}, one has
\be
k_{t}=\alpha, \;\;\;\;\;\;\;\;\; k_{z}=-\alpha\frac{\partial t_{+}}{\partial z}, \;\;\;\;\;\;\;\;\;  k_{a}=-\alpha\frac{\partial t_{+}}{\partial \sigma^a}.
\ee
It is straightforward to show 
\bea
k^{z} &=&-\alpha G^{zz}\frac{\partial t_{+}}{\partial z}=\frac{-\alpha z^2}{L^2}(1+\frac{z^2}{2}g_{tt}^{(1)}+...),
\nonumber
\\
k^{t}&=&\alpha G^{tt}-\alpha G^{ta}\frac{\partial t_{+}}{\partial \sigma^a},
\nonumber
\\
k^{a}&=&\alpha G^{at}-\alpha G^{ab}\frac{\partial t_{+}}{\partial \sigma^a}.
\label{k1-up}
\eea
On the other hand, from 
\be
G_{ij}=\frac{L^2}{z^2}(g^{(0)}_{ij}+z^2g^{(1)}_{ij}+...) ,\,\,\,\,\,g^{(0)}_{tt}=-1,\,\,\,\,\,\, g^{(0)}_{ta}=0,
\ee
and
\be
 g^{(0)}_{ab}(t,\sigma)=h_{ab}+t\partial_{t} g^{(0)}_{ab}+\frac{t^2}{2}\partial_{t}^2g^{(0)}_{ab}+...
\ee
one can write the inverse metric on the null surface as follows
\bea
G^{tt}&=&\frac{-z^2}{L^2}(1+z^2g_{tt}^{(1)}+\cdots),
\nonumber
\\
G^{ta}&=&\frac{z^4}{L^2}g_{tb}^{(1)}h^{ba}+\cdots
\nonumber
\\
G^{ab}&=& \frac{z^2}{L^2}h^{ab}+\cdots.
\eea
Then from eq. \eqref{k1-up} to second order in $z$ we have
\bea
k^t=\frac{-\alpha z^2}{L^2}(1+z^2g_{tt}^{(1)}+...), \;\;\;\;\;\;\;\;\;\;\;\;\;\;
k^a=\frac{\alpha z^4}{L^2}g_{tb}^{(1)}h^{ba}.
\label{k1-down}
\eea
Putting everything together, eq. \eqref{affine-coord-1} leads to 
\bea
\frac{\partial z(\lambda,\rho)}{\partial\lambda} &=&\frac{-\alpha z^2}{L^2}(1+\frac{z^2}{2}g_{tt}^{(1)}(\sigma)+\cdots),
\nonumber\\
\frac{\partial t(\lambda,\rho)}{\partial\lambda} &=&\frac{-\alpha z^2}{L^2}(1+z^2g_{tt}^{(1)}(\sigma)+\cdots),
\nonumber\\
\frac{\partial \sigma^a(\lambda,\rho)}{\partial\lambda}&=&\frac{\alpha z^4}{L^2}g_{tb}^{(1)}h^{ba}+\cdots.
\label{affine-eq}
\eea
In the third equation, the first term is at order $\mathcal{O}(z^4)$, and hence one can conclude that to first order in $z$$, \sigma$ is independent of $\lambda$. Next, one can take $\sigma$ a constant and solve the first equation in eq. \eqref{affine-eq}. To second order in $z$, one can easily find
\be
\lambda=\frac{L^2}{\alpha z} \left( 1+\frac{g_{tt}^{(1)}}{2}z^2+\cdots \right).
\label{Lambda-affine-future}
\ee
\section{Holographic complexity of $AdS_{d+1} $ in global coordinates}
\label{App-B}
In this section we calculate the holographic complexity of a $AdS_{d+1}$ spacetime in global coordinates for $d=2,3,4$. It should be pointed out, this subject has been partly addressed in ref. \cite{Chapman:2016hwi}. Moreover, the structure of its UV divergences has been studied in refs. \cite{Carmi:2016wjl,Reynolds:2016rvl}.
\footnote{Recently, the holographic complexity of $AdS_3$ in Poincar\'e coordinates and it changes under infinitesimal conformal transformations in the dual CFT have been studied in ref. \cite{Flory:2019kah}.}
In global coordinates the corresponding metric is given by 
\bea
ds^2 =\frac{L^2}{r^2} \left(-f(r) dt^2+ \frac{1}{f(r)} dr^2 +L^2 d\Omega^2_{d-1}\right),\;\;\;\;\;
\;\;\;
f(r)= 1+ \frac{r^2}{L^2}.
\label{metric-AdS}
\eea 
The tortoise coordinate can be calculated easily
\bea
r^*(r) =- \int \frac{dr}{f(r)} =- L \arctan \frac{r}{L} + c,
\eea 
here, $c$ is the constant of integration. Therefore, one has
\bea
r^*(\delta)- r^*(r) = L \arctan \frac{r}{L}.
\eea 
Recall that we work with the second regularization. The bulk action is given by
\bea
I_{\rm bulk} = - \frac{d V_{d-1} L^{2(d-1)}}{2\pi G_N} \int_{\delta}^{r_{\rm max}} \frac{dr}{r^{d+1}} \left(r^*(\delta) - r^*(r) \right).
\eea 
The contribution of the null-null joints at the tip and bottom of the WDW are easily zero \cite{Chapman:2016hwi} (See appendix B in ref. \cite{Chapman:2016hwi}). The only remaining joints are a null-null one at the UV cutoff surface, i.e. $r=\delta$, and its contribution to the action is given by 
\bea
I_{\rm joints} = - \frac{V_{d-1} L^{2(d-1)}}{8\pi G_N} \frac{1}{\delta^{d-1}} \left[ \log\frac{\alpha \beta \delta^2}{L^2} - \log |f(\delta) |\right].
\eea 
The normal vectors to the future null boundary $B'_{1}$ and the past null boundary $B'_{2}$ are given by
\bea
k'_1 = \alpha \left(dt,-\frac{dr}{f(r)}\right) , \;\;\;\;\;\;\;\;\;\;\;\;\;\;\;\;\;\;  k'_2 = \beta \left(dt,\frac{dr}{f(r)}\right).
\eea 
The contributions of the two null surfaces in the counterterm $I_{\rm ct}^{(0)}$ are given by
\bea
I_{\rm ct}^{(0)} &=&- \frac{V_{d-1} L^{2(d-1)}}{8\pi G_N} \bigg[\frac{1}{r_{\rm max}^{d-1}} \left( \frac{2}{(d-1)}+ \log\frac{\alpha \beta r_{\rm max}^2}{L^2}\right)
-\frac{1}{\delta^{d-1}} \left( \frac{2}{(d-1)}+ \log\frac{\alpha \beta \delta^2}{L^2}\right) \bigg]
\cr&& \cr&&
\!\!\!\!= \frac{V_{d-1} L^{2(d-1)}}{8\pi G_N} \frac{1}{\delta^{d-1}} \left[\frac{2}{(d-1)}+ \log\frac{\alpha \beta \delta^2}{L^2}\right].
\eea 
In the last line, we took the limit $r_{\rm max} \rightarrow \infty$.
\\It is straightforward to show that for global AdS eq. \eqref{Theta term-Null counterterm} is satisfied, and hence for each null boundary the counterterm $I_{\rm ct}^{(1)}$, i.e. eq. \eqref{I-ct-1} is reduced to eq. \eqref{I-ct-1-d}. Therefore, it is easy to show for the two null boundaries one has
\begin{itemize}
	\item for $d=3$
	\bea
	I_{\rm ct}^{(1)} = - \frac{V_{d-1} L^2}{8 \pi G_N} \log \frac{r_{\rm max}}{\delta},
	\eea
	\item for $d \neq 3$
	\bea
	I_{\rm ct}^{(1)} &=&  -\frac{V_{d-1} (d-1)L^{2(d-2)}}{16 \pi G_N (d-3)}\left(  \frac{1}{\delta^{d-3}} -  \frac{1}{r_{\rm max}^{d-3}} \right)
	\cr&& \cr&& 
	= -\frac{V_{d-1} (d-1)L^{2(d-2)}}{16 \pi G_N (d-3)} \frac{1}{\delta^{d-3}},
	\eea
\end{itemize}
Putting everything together, one has
\begin{itemize}
\item for $d=2$, the counterterm $I_{\rm ct}^{(1)}=0$, and one has
\bea
\mathcal{C}_{{\rm AdS}_{3}}= \frac{V_1 L}{8 \pi G_N} =  \frac{ L}{4 G_N}.
\label{complexity-AdS-d=2}
\eea 
This is exactly the finite part of the holographic complexity obtained in eq. (4.5) in ref. \cite{Chapman:2016hwi}.
\item for $d=3$
\bea
\mathcal{C}_{{\rm AdS}_{4}}= \frac{V_2 L^2}{8 \pi^2 G_N} \left( 1 + 2 \log \frac{L}{r_{\rm max}}\right).
\label{complexity-AdS-d=3}
\eea 
\item for $d=4$
\bea
\mathcal{C}_{{\rm AdS}_{5}}=- \frac{V_3 L^3}{8 \pi G_N}.
\label{complexity-AdS-d=4}
\eea 
\end{itemize}
Therefore, the null counterterms eq. \eqref{I-ct-1} remove all the UV divergences of the holographic complexity of global AdS. However, the structure of them is such that they introduce a logarithmic IR divergent term $\log r_{\rm max}$ for odd d. As discussed in section \ref{sec: New Counterterm on the Singularity}, for AdS-Schwarzschild black holes we are able to remove the term by adding the counterterm eq. \eqref{I-ct-2}. In contrast, for a global AdS there is not such a possibility.



\end{document}